\documentclass{article} 
\usepackage[dvipsnames]{xcolor}   %
\usepackage{iclr2026_conference,times}

\usepackage{amsmath,amsfonts,bm}

\def\eqref#1{equation~\ref{#1}}

\def\1{\bm{1}}

\def\rc{{\textnormal{c}}}

\def\rx{{\textnormal{x}}}
\def\ry{{\textnormal{y}}}
\def\rz{{\textnormal{z}}}

\def\rvu{{\mathbf{i}}}

\def\rvu{{\mathbf{u}}}

\def\rvx{{\mathbf{x}}}
\def\rvy{{\mathbf{y}}}
\def\rvz{{\mathbf{z}}}

\def\vc{{\bm{c}}}

\def\vr{{\bm{r}}}

\def\mD{{\bm{D}}}

\def\mF{{\bm{F}}}

\def\mI{{\bm{I}}}
\def\mJ{{\bm{J}}}

\def\mP{{\bm{P}}}

\def\mR{{\bm{R}}}

\def\mT{{\bm{T}}}

\def\mW{{\bm{W}}}
\def\mX{{\bm{X}}}

\DeclareMathAlphabet{\mathsfit}{\encodingdefault}{\sfdefault}{m}{sl}
\SetMathAlphabet{\mathsfit}{bold}{\encodingdefault}{\sfdefault}{bx}{n}

\usepackage{hyperref}
\usepackage{url}

\usepackage{algorithm}
\usepackage{algpseudocode}
\usepackage{array}
\usepackage{booktabs}
\usepackage{multirow}
\usepackage{dsfont}
\usepackage{color}
\usepackage{colortbl}
\usepackage{graphicx}
\usepackage{booktabs}
\usepackage{amsmath}
\usepackage{amssymb}
\usepackage{mathtools}
\usepackage{amsthm}
\usepackage{multirow}
\usepackage{booktabs}
\usepackage{subcaption}
\usepackage{relsize}
\usepackage{titletoc} 
\usepackage{tikz}
\usepackage{comment}

\usepackage[dvipsnames]{xcolor}   %
\usepackage{pifont}    %
\usepackage[percent]{overpic}

\usepackage[capitalize,noabbrev]{cleveref}
\usepackage{float}
\usepackage{float}
\usepackage[inkscapelatex=false]{svg}
\usepackage{courier}

\makeatletter
\AddToHook{cmd/appendix/before}{\def\cref@section@alias{appendix}}
\makeatother

\crefname{section}{Sec.}{Secs.}
\Crefname{section}{Section}{Sections}
\Crefname{table}{Table}{Tables}
\crefname{table}{Tab.}{Tabs.}
\Crefname{append}{Appendix}{Appendixs}
\crefname{append}{Append.}{Appends.}
\Crefname{subfigure}{Figure}{Figures}
\crefname{subfigure}{Fig.}{Figs.}

\definecolor{ForestGreen}{RGB}{34,139,34}
\definecolor{myyellow}{RGB}{181, 181, 27}
\newcommand{\cmark}{\ding{51}}%
\newcommand{\xmark}{\ding{55}}%
\newcommand{\greencheck}{{\color{ForestGreen}\cmark}}
\newcommand{\yellowcheck}{{\color{myyellow}\cmark}}
\newcommand{\redcheck}{{\color{red}\xmark}}

\definecolor{tabfirst}{rgb}{1, 0.7, 0.7} 
\definecolor{tabsecond}{rgb}{1, 0.85, 0.7} 
\definecolor{tabthird}{rgb}{1, 1, 0.7} 

\newcommand{\xy}{\begin{bmatrix} \rvx \\ \rvy \end{bmatrix}}
\newcommand{\xyt}{\left[ \rvx_t, \rvy_t \right]^\top}
\newcommand{\xytc}{\left[ \rvx_t^c, \rvy_t^c \right]^\top}
\newcommand{\ee}{\bm{\varepsilon}}
\newcommand{\mm}{\bm{\mu}}

\newcommand{\ind}[1]{\mathds{1}_{#1}}

\definecolor{rowcolor}{HTML}{ABB5E9}
\definecolor{teaserblue}{HTML}{55A8D6}
\definecolor{teaserpurple}{HTML}{926AC9}

\newcommand{\newparallel}{/\!\!/}

\theoremstyle{plain}
\newtheorem{theorem}{Theorem}[section]

\newtheorem{lemma}[theorem]{Lemma}

\theoremstyle{definition}

\theoremstyle{remark}

\newcommand{\dataname}{PulpMotion}
\newcommand{\ours}{Aux}

\newcommand{\director}{$(\rvx)$+{\sc Director}}
\newcommand{\indxy}{$(\rvx)$$(\rvy)$}
\newcommand{\jxy}{$(\rvx, \rvy)$}
\newcommand{\jxyz}{$(\rvx, \rvy, \rvz)$}
\newcommand{\ourxy}{$(\rvx, \rvy)$+\ours{}}
\newcommand{\ourindxy}{$(\rvx)(\rvy)$+\ours{}}

\title{Pulp Motion: Framing-aware multimodal \\
camera and human motion generation
}

\author{Robin Courant$^{1}$
\quad
Xi Wang$^{1}$
\quad
David Loiseaux$^{1,2}$
\quad
Marc Christie$^{3}$
\quad
Vicky Kalogeiton$^{1}$\\
$^1$LIX, École Polytechnique, CNRS, IPP \quad
$^2$Inria Saclay \quad
$^3$Inria, IRISA, Univ Rennes, CNRS}

\iclrfinalcopy 
\begin{document}

\maketitle

\begin{abstract}
Treating human motion and camera trajectory generation separately overlooks a core principle of cinematography: the tight interplay between actor performance and camera work in the screen space. 
In this paper, we are the first to cast this task as a text-conditioned joint generation, aiming to maintain consistent on-screen framing while producing two heterogeneous, yet intrinsically linked, modalities: human motion and camera trajectories. 
We propose a simple, model-agnostic framework that enforces multimodal coherence via an auxiliary modality: the on-screen framing induced by projecting human joints onto the camera. This on-screen framing provides a natural and effective bridge between modalities, promoting consistency and leading to more precise joint distribution.
We first design a joint autoencoder that learns a shared latent space, together with a lightweight linear transform from the human and camera latents to a framing latent. We then introduce auxiliary sampling, which exploits this linear transform to steer generation toward a coherent framing modality. 
To support this task, we also introduce the PulpMotion dataset, a human-motion and camera-trajectory dataset with rich captions, and high-quality human motions.
Extensive experiments across DiT- and MAR-based architectures show the generality and effectiveness of our method in generating on-frame coherent human-camera motions, while also achieving gains on textual alignment for both modalities. Our qualitative results yield more cinematographically meaningful framings setting the new state of the art for this task.
Code, models and data are available in our \href{https://www.lix.polytechnique.fr/vista/projects/2025_pulpmotion_courant/}{project page}.
\end{abstract}

\section{Introduction}
\label{sec:introduction}
Cinematography is inherently a collaborative task, shaped by the joint relationship between the actor and the director.
On the one hand, the director’s camera seeks to frame the actors, adjusting to their movements to capture the desired performance on screen. 
On the other hand, the actor must also remain attentive to the presence of the camera, \emph{e.g.} pausing at a marker until the camera arrives, before continuing a movement. 
Such motions are not spontaneous but rather intentional. These carefully crafted choices aim at enhancing the cinematic aesthetics.
Balancing natural performance with visual framing between actors and cameras remains a central challenge in filmmaking.

Prior works have typically addressed only one side of this joint problem, treating them as standalone modalities: either human motion generation~\citep{zhang2024motiondiffuse,tevet2022mdm,jiang2024motiongpt} or camera trajectory generation~\citep{jiang2024ccd,courant2024exceptional,zhang2025gendop}, but never both simultaneously.
In this work, we introduce the text-conditioned task of jointly generating human motion and camera trajectories. This task is challenging, as any mismatch between motion and camera may lead to poor framing, how the characters are positioned on screen, or even empty frames (\emph{e.g.}, the subject moving out of view).
The root problem of this joint generative task, referred to in computer vision as multimodal generation, is to produce high-quality outputs for each modality while maintaining multimodal coherence. 

Multimodal generation has been widely studied in domains such as video–audio~\citep{ruan2023mm,hayakawa2024mmdisco} and image–text~\citep{li2025dual,xu2023versatile}. However, most approaches either rely solely on paired data to capture multimodal relationships~\citep{xieshow,li2025dual,swerdlow2025unified}, explicitly enforce correlations through architectural or algorithmic designs~\citep{hu2022unified,ruan2023mm,xu2023versatile,tang2023codi}, or require training adaptations or sampling guided by models trained on external data~\citep{bao2023one,hayakawa2024mmdisco,xing2024seeing,kouzelis2025boosting}.

Training only on paired data provides an incomplete approximation of the joint data distribution (often due to mode coverage), making it challenging to sample precisely coherent modality pairs during generation. To address this, rather than adding architectural complexity as other methods do, we propose a multimodal generation framework that leverages an auxiliary modality as a bridge between generated modalities, steering the sampling process toward regions of higher multimodal coherence.
Concretely, as illustrated in~\Cref{fig:teaser}, the model approximates an imperfect joint human–camera distribution (shown in blue). To mitigate this, we leverage the on-screen human framing within the camera as an auxiliary modality to steer the sampling toward a more coherent region of the joint distribution (shown in purple), \emph{i.e.}, human and camera pair with a cinematic framing.

%
Our framework consists of two stages:
(1) learning a joint latent space for human motion and camera trajectories, along with a linear transform which maps them into the auxiliary modality, \emph{i.e.} the on-screen framing. This linear transform captures the relationship between the generated and auxiliary modalities directly in the latent space;
(2) a sampling process augmented with an additional term derived from this linear transform, steering generation towards coherent multimodal generation.

For evaluation, we present \dataname{}, an extended version of a prior human-camera dataset, with more samples, motion captions, and higher-quality motion. 
We benchmark our approach on this dataset for both DiT-based~\citep{peebles2023scalable} and MAR-based~\citep{li2024mar} architectures to demonstrate the generality and model-agnosticity of our approach. 
Our results show consistently improved coherence between generated motion and trajectories, yielding better framing quality and lower out-of-frame rates while preserving strong motion and trajectory generation performance.

Our contributions are: 
(1) a unified framework that jointly generates human motion and camera trajectories leveraging an auxiliary modality (on-screen framing) to enforce multimodal coherence during sampling,
(2) the \dataname{} dataset, an extension of the prior human-camera dataset with  more samples, motion captions, and higher-quality human motion, and 
(3) an extensive evaluation across multiple architectures demonstrating the method’s generality and effectiveness.

\begin{figure*}[t]
    \centering
    \input{fig/teaser.tex}
    \label{fig:teaser}
\end{figure*}

\section{Related work}
\label{sec:related-work}
\noindent
\textbf{Human motion generation.} Diffusion-based approaches have driven recent progress~\citep{ho2020ddpm,rombach2022ldm,tevet2022mdm,kim2023flame,zhang2024motiondiffuse} on human motion generation, with extensions for efficient latent spaces, fast sampling, stronger textual alignment, and leverage of external data~\citep{chen2023mdl,dai2024motionlcm,andreou2025lead,zhang2023remodiffuse}.
The newly proposed MAR architecture combines autoregressive and diffusion modeling and further pushes state-of-the-art performance~\citep{li2024mar,meng2024mardm,xiao2025motionstreamer}.
\newline However, most methods treat motion as a \emph{single modality}; although interactions with objects, people, and scenes are increasingly modeled~\citep{peng2025hoi,geng2025auto,liang2024intergen,fan2024freemotion,shan2024towards,wang2024move,cen2024generating}, joint human–camera generation remains largely unexplored. Existing efforts typically use camera parameters only as constraints or conditioning, rather than modeling their joint distribution with motion~\citep{Priyanka20253dv,ye2023slahmr,wang2024tram,kocabas2024pace,Sun_2023_CVPR}.


\noindent
\textbf{Camera Trajectory Generation.} Camera control has evolved from handcrafted rules to learning-based methods that either mimic cinematic from example videos or optimize trajectories in differentiable 3D space~\citep{blinn1988looking,lino2015intuitive,drucker1992cinema,jiang20sig,jiang21siga,wang2023jaws,jiang2024cinematic,chen2024dreamcinema}. 
To reduce reliance on exemplary data, Reinforcement Learning (RL)-based methods are often applied on drones and indoor scenes~\citep{Huang_2019_CVPR,bonatti2020autonomous,Xie_2023_ICCV}, but they remain environment-specific and style-limited. 
%
%
Diffusion-based camera generation, coupled with new datasets, further advances text-conditioned control and reduces reliance on curated reference videos~\citep{jiang2024ccd, courant2024exceptional, wang2024dancecamanimator,wang2024dancecamera3d, zhang2025gendop}.
\newline However, similarly to human motion generation, camera generation is also often regarded as a \emph{single modality} problem conditioned on motion, rather than modeling the joint motion–camera distribution. In this work, we bridge this gap by adding human motion into the camera trajectory generation pipeline, modelling the synergy between how and what to film.

\noindent
\textbf{Multimodal generation.} Most multimodal generation works leverage paired data to implicitly capture joint distribution, \emph{e.g.}, text–image unified generation has been explored with different architectures: Dual Diffusion~\citep{li2025dual} employs a DiT-based design, while Show-o~\citep{xieshow} adopts an autoregressive backbone plus diffusion head framework.
However, in practice, relying solely on paired data to learn implicit multimodal correlations often requires large datasets and still fails to fully capture multi-modal relationships.

Therefore, some works explicitly enforce multimodal coherence through architectural or algorithmic design.
~\cite{hu2022unified} introduce a unified transition that compresses discrete representations across modalities under a discrete diffusion framework. 
MMDiffusion~\citep{ruan2023mm} exploits a similar idea. It employs multimodal attention and random shifts to align multimodal information. \cite{xu2023versatile} emphasizes architectural separation by disentangling context and data layers to encourage joint conditioning. 
Alternatively, CoDi~\citep{tang2023codi} modifies cross-attention layers to emphasize a pre-aligned, modality-specific latent space, enabling any-to-any generation across multiple modalities.
Despite their effectiveness in specific tasks, these approaches often depend on hand-crafted architectures, which limit their generality and adaptability across tasks and models.

%
Another line of work focuses on adapting only the training or sampling process, avoiding architectural modifications.
For instance, UniDiffuser~\citep{bao2023one} trains a single multimodal diffusion with independent timesteps for each modality and applies an adapted classifier-free guidance scheme (CFG)~\citep{ho2021cfg} with modality-specific timesteps.
MMDisco~\citep{hayakawa2024mmdisco}, inspired by classifier guidance~\citep{dhariwal2021diffusion}, enables video–audio generation by training a joint discriminator to construct a guidance term during sampling, which is also used as regularisation for fine-tuning. 
Meanwhile, some works leverage foundation models to implicitly exploit larger datasets and stronger representations. For example, \cite{xing2024seeing} use the pre-trained multimodal binder ImageBind~\citep{girdhar2023imagebind} to align generations via classifier-like guidance. Similarly, \cite{kouzelis2025boosting} design a representation-guidance term based on a diffusion generator trained on paired DINOv2~\citep{oquab2024dinov2} and image data, preserving the joint distribution without requiring an explicit classifier.
However, these approaches still rely on adapting training or using large external pre-trained models, such as Imagebind or DINOv2.

In this work, we leverage an auxiliary modality to bridge the target modalities, steering sampling toward coherent joint generation in an architecture-agnostic manner, without requiring training adaptations or pre-trained models on external data. See extended discussion in Appendix~\ref{supp:related work}.

\section{Method}
\label{sec:method}
\noindent
\textbf{Problem formulation.}
We consider a sample as a pair of human motion $\rvx \in \mathbb{R}^{N_\rvx}$ and camera trajectory $\rvy \in \mathbb{R}^{N_\rvy}$; both are sequences of $F$ frames.
We aim to generate both modalities with respect to a textual description $\rc \in \mathbb{R}^{N_\rc}$, specifying the desired human motion and camera trajectory, \emph{i.e.}, sampling from the joint human-camera distribution $p(\rvx, \rvy | \rc)$.

\noindent
\textbf{Our approach.} 
Most related works capture multimodal relationships by relying exclusively on paired data, crafting specific architectural and algorithmic designs, introducing training adaptations or sampling strategies guided by external models.
In contrast, our approach strengthens the coherence between modalities using an auxiliary modality, $\rvz  \in \mathbb{R}^{N_\rvz}$ (with $N_\rvz < N_\rvx+N_\rvy$), which explicitly bridges them.
In our setting, $\rvz$ represents the on-screen human framing, \emph{i.e.}, the 2D projection of human joints in the camera view, a natural characteristic of the human-camera relationship.

Next, we describe our multimodal latent space with a latent linear transform of the auxiliary modality (Section~\ref{sec:latent-space}) and present our auxiliary sampling scheme, which leverages the relationship between generated modalities and the auxiliary modality  (Section~\ref{sec:diffusion-framework}).


\subsection{Multimodal latent space}
\label{sec:latent-space}

In multimodal generation, different modalities often exhibit varying properties, such as scale or geometric structure, which makes direct generation in the raw modality space challenging~\citep{tang2023codi,xing2024seeing}. 
Moreover, operating directly in the raw space increases computational and memory costs~\citep{rombach2022ldm} and can destabilize some diffusion losses~\citep{meng2024mardm}.

To address these challenges, we adopt a latent diffusion framework for our joint human-camera generation task. Our latent representation is designed with two key aspects: (1) instead of embedding modalities separately, human and camera are aligned into a shared latent space, and (2) a lightweight learnable linear transform $\mW \in \mathbb{R}^{(N_\rvx+N_\rvy)\times N_\rvz}$ that maps the latent human and camera representations into an on-screen framing latent, bridging both modalities. 

\begin{figure}[t]
\centering
\begin{minipage}{0.53\textwidth}
  \centering    
  \vspace{12pt}
  \input{fig/autoencoder/autoencoder}
  \captionof{figure}{\small \textbf{Architecture of the multimodal autoencoder.} Human motion $\rvx_{\text{raw}}$ and camera trajectory $\rvy_{\text{raw}}$ are jointly encoded by $E_\phi$, linearly transformed via $\mW$ into an auxiliary on-screen framing latent $\rvz$. Three decoders $D_{\psi_\rx}$, $D_{\psi_\ry}$, and $D_{\psi_\rz}$ reconstruct raw modalities: $\hat{\rvx}_{\text{raw}}, \hat{\rvy}_{\text{raw}}, \hat{\rvz}_{\text{raw}}$.}
  \label{fig:autoencoder}
\end{minipage}%
\hspace{8pt}
\begin{minipage}{0.39\textwidth}
  \centering
  \includegraphics[width=\textwidth]{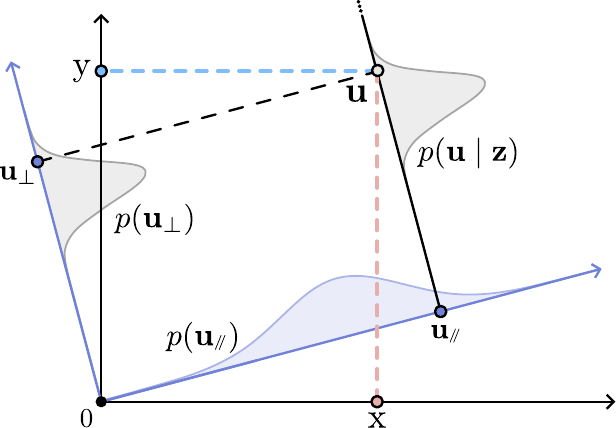}
    \captionof{figure}{\small \textbf{Decomposition of $\rvu =
    [\rvx,\rvy]^\top$.} $\rvu$ decomposes onto two orthogonal components
    $\rvu_\perp$ and $\rvu_{\newparallel}$. Our auxiliary sampling leverages
    this to encourage samples along $\rvu_{\newparallel}$, parallel to the
    auxiliary modality $\rvz$.}
  \label{fig:xyz_rpz}
\end{minipage}
\end{figure}

More specifically, we propose an autoencoder architecture, shown in Figure~\ref{fig:autoencoder}. The model first maps the human $\rvx_{\text{raw}}$ and camera $\rvy_{\text{raw}}$ with a joint encoder $E_{\phi}$, producing latent embeddings $\rvx$ and $\rvy$. 
A learnable linear transform $\mW$ then maps these embeddings into a on-screen framing latent $\rvz$:
\begin{align}
    \rvz = \mW \left[ \rvx, \rvy \right]^\top \quad . 
    \label{eq:latent-proj}
\end{align}
Finally, three \textit{independent} decoders $D_{\psi_\rx}$, $D_{\psi_\ry}$ and $D_{\psi_\rz}$ reconstruct each raw modality ($\rvx_{\text{raw}}, \rvy_{\text{raw}}, \rvz_{\text{raw}}$) from its respective latent\footnote{Recall $\rvz_{\text{raw}}$ is defined as the 2D projection of human joints in the camera view, see \Cref{sec:xp-setup} for details for $\rvx_{\text{raw}}, \rvy_{\text{raw}}, \rvz_{\text{raw}}$.}. The model is trained end-to-end with the following loss:
\begin{equation}
    \begin{split}
        \mathcal{L}_{\text{AE}}(\phi, \psi_c, \psi_h, \psi_p) & = \Vert D_{\psi_\rx}(E_{\phi}(\rvx_{\text{raw}}, \rvy_{\text{raw}})) - \rvx_{\text{raw}} \Vert^2  
        + \Vert D_{\psi_\ry}(E_{\phi}(\rvx_{\text{raw}}, \rvy_{\text{raw}})) - \rvy_{\text{raw}} \Vert^2       \\
      & \quad + \Vert D_{\psi_\rz}(\mW E_{\phi}(\rvx_{\text{raw}}, \rvy_{\text{raw}})) - \rvz_{\text{raw}} \Vert^2 \quad .
        \label{eq:ae-loss}
    \end{split}
\end{equation}
Note that the on-screen framing is never directly encoded; it is learned exclusively via the linear transform from the camera and human latents and supervised only through its reconstruction loss.

\subsection{Auxiliary sampling}
\label{sec:diffusion-framework}

Given the multimodal latent space established in the previous section, we now introduce our multimodal latent diffusion framework, which incorporates an auxiliary sampling technique during inference to enhance multimodal coherence.

We train a generative model to produce multimodal representations of human motion $\rvx$ and camera trajectories $\rvy$ from textual descriptions $\rc$. For this, we adopt the standard Denoising Diffusion Probabilistic Model (DDPM) framework~\citep{ho2020ddpm}:
\begin{align}
    \mathcal{L}_{\text{noise}}(\theta) = \mathbb{E}_{t,\ee_{\rvx\rvy}} \left[ \Vert \ee_{\rvx\rvy} - \ee_\theta \left(\rvx_t, \rvy_t, \rc \right) \Vert^2 \right] \quad ,
\end{align}
where $\ee_\theta$ denotes the model’s predicted noise corresponding to the true noise $\ee_{\rvx\rvy}$ at timestep $t$.

\paragraph{Auxiliary sampling.}
Controllability in diffusion models is typically achieved via classifier-free guidance (CFG)~\citep{ho2021cfg} over a conditioning signal $\vc$: 
\begin{equation}
    \begin{split}
        \nabla_{\rvx_t, \rvy_t} \log \tilde{p}(\rvx_t, \rvy_t | \rc)
         & = \nabla_{\rvx_t, \rvy_t} \log p(\rvx_t, \rvy_t )         \\
         & \quad +  w_c (\nabla_{\rvx_t, \rvy_t} \log p(\rvx_t, \rvy_t | \rc) - \nabla_{\rvx_t, \rvy_t} \log p(\rvx_t, \rvy_t )) \quad .
    \label{eq:cfg}
    \end{split}
\end{equation}
Note that the model’s output is proportional to the score: $\ee_\theta(\rvx) \propto \nabla_\rvx \log p(\rvx)$.

Here, CFG explicitly splits the score into an unconditional term and a conditional term, with the latter scaled by $w_c$.
In our case, we aim to control the multimodal coherence between $\rvx$ and $\rvy$ through $\rvz$. Following the CFG strategy, we split the unconditional score term $\nabla_{\rvx_t, \rvy_t} \log p(\rvx_t, \rvy_t)$ in \Cref{eq:cfg} into a new “unconditional” term and an additional term in $\rvz$. To this end, we leverage the relationship in \Cref{eq:latent-proj} that links $\rvz$ to $\rvx$ and $\rvy$.
\newline Let $\rvu = [\rvx, \rvy]^\top \in \mathbb{R}^{N_\rvx+N_\rvy}$. Since $\rvz$ is a compressed representation of $\rvu$ (\emph{i.e.}, $N_\rvz < N_\rvx+N_\rvy$), it cannot fully capture all information in $\rvu$.
We therefore decompose $\rvu$ into a $\rvz$-dependent component $\rvu_{\newparallel}$ (\emph{i.e.}, $\rvu_{\newparallel} \mapsto \rvz = \mW\rvu_{\newparallel}$ is an isomorphism) and a complementary orthogonal component $\rvu_{\perp}$, such that $\rvu = \rvu_{\perp} + \rvu_{\newparallel}$, as illustrated in \Cref{fig:xyz_rpz}.
This decomposition is precisely what we aim for: the component
$\rvu_{\newparallel}$ characterized by $\rvz$
, steers the sampling
toward a coherent $\rvu$, while the complementary component acts as an
“unconditional” term.

\begin{lemma}
Let $\mP_{\newparallel}$ denote the projection onto the orthogonal space of $\ker (\mW)$.  
Then, we have:
\begin{equation}\label{eq:p-distributions}
    \rvu_\perp := (\mI - \mP_{\newparallel}) \rvu \sim \mathcal{N}((\mI - \mP_{\newparallel}) \mm, \sigma^2 (\mI - \mP_{\newparallel}))
    \quad \text{and} \quad
    \rvu_{\newparallel} := \mP_{\newparallel} \rvu \sim \mathcal{N}(\mP_{\newparallel} \mm, \sigma^2 \mP_{\newparallel}) \quad ,
\end{equation}
and the density of $\rvu$ decomposes as:
\begin{equation}\label{eq:decomposition-lemma}
    p(\rvu) = p(\rvu_\perp) \, p (\rvu_{\newparallel}).
\end{equation}
\end{lemma}
Since $\mW^\top \mW$ is invertible in our setting, $\mP_{\newparallel}$ can be expressed as $\mP_{\newparallel} = \mW (\mW^\top \mW)^{-1} \mW^\top$.
See \Cref{fig:xyz_rpz} for illustration and Appendix~\ref{sec:theory:method} for complete development and proof.

Thus, using \Cref{eq:decomposition-lemma}, the first term in \Cref{eq:cfg} can be split into an “unconditional” term over $(\rvx_t, \rvy_t)$ and a $\rvz$-dependent term weighted by $w_z$ that steers sampling toward $\rvz_t$:
\begin{equation}
    \begin{split}
        \nabla_{\rvx_t, \rvy_t} \log \tilde{p}(\rvx_t, \rvy_t| \rc)
         & = \nabla_{\rvx_t, \rvy_t} \log p(\rvu_{\perp}) 
         + (1 + w_z) \nabla_{\rvx_t, \rvy_t} \log p(\rvu_{\newparallel})                \\
         & \quad +  w_c (\nabla_{\rvx_t, \rvy_t} \log p(\rvx_t, \rvy_t | \rc)
        - \nabla_{\rvx_t, \rvy_t} \log p(\rvx_t, \rvy_t)) \quad .
    \end{split}
    \label{eq:cfg-z}
\end{equation}
Finally, we perform sampling using the following linear combination of the model's predictions, recalling that $\ee(\rvx) \propto \nabla_\rvx \log p(\rvx)$, with more detailed derivation included in Appendix~\ref{sec:aux-derivation}:
\begin{equation}
    \begin{split}
        \ee_\theta \left( \rvx_t, \rvy_t, \rc, t \right)
         & = \ee_\theta \left( \rvx_t, \rvy_t, \emptyset, t \right)
         + w_z \mP_{\newparallel} \; \ee_\theta\left(\rvx_t, \rvy_t, \emptyset, t \right) \\
         & \quad + w_c (\ee_\theta \left( \rvx_t, \rvy_t, \rc, t \right)  - \ee_\theta \left( \rvx_t, \rvy_t, \emptyset, t \right) )  \quad .
    \end{split}
    \label{eq:epsilon-cfg-z}
\end{equation}
Since the final sampling does not explicitly condition on $\rvz_t$, there is no need to include the auxiliary modality during training. This reduces training cost and yields a more general approach.

\section{\dataname{} dataset}
\label{sec:dataset}
\begin{table}[t]
\caption{\small \textbf{Comparison of human and camera datasets.} We compare \dataname{} with existing human motion and/or camera trajectory datasets. We summarize modality coverage, available captions, dataset size (hours, frames, samples), sample length statistics (median, mean, std), and vocabulary size.}
\label{tab:datasets}
\centering
\resizebox{\textwidth}{!}{%
\begin{tabular}{lc@{\hspace{3mm}}c@{\hspace{2mm}}c@{\hspace{1mm}}crrrrrrr}
\toprule
\multirow{2}{*}{\textbf{Dataset}} & \multicolumn{2}{c}{\textbf{Camera}} & \multicolumn{2}{c}{\textbf{Human}} & \multirow{2}{*}{\textbf{\#Hours}} & \multirow{2}{*}{\textbf{\#Frames}} & \multirow{2}{*}{\textbf{\#Samples}} & \multicolumn{3}{c}{\textbf{Sample lengths (frames)}} & \multirow{2}{*}{\textbf{\#Vocabulary}} \\
       & Traj & Caption & Motion & Caption &    &         &          & Median      & Mean      & Std      &  \\
\midrule
RealEstate10k~\citep{zhou2018realestate10k} & \greencheck & \redcheck & \redcheck & \redcheck & 121 & 11M & 79K & 115 & 136.9 & 80.0 & -   \\
CamVid-30K~\citep{zhao2024camvid30k}  & \greencheck & \redcheck & \redcheck & \redcheck   & - & - & 30K & - & - & - & -  \\
DynPose100k~\citep{rockwell2025dynpose100k}  & \greencheck & \redcheck & \redcheck & \redcheck  & 157 & 6.8M & 100K & 63 & 67.97 & 17.91 & -   \\
CameraBench~\citep{lin2025camerabench}  & \redcheck & \greencheck & \redcheck & \redcheck  & - & - &  4K & - & - & - & -   \\
CCD~\citep{jiang2024ccd}  & \greencheck & \greencheck & \redcheck & \redcheck  & 50 & 4.5M & 25K & 189 & 180.4 & 69.6 & 48   \\
DataDoP~\citep{zhang2025gendop}  & \greencheck & \greencheck & \redcheck & \redcheck  & 113 & 11M & 29K & - & 424.8 & - & 8,698   \\
\midrule
KIT-ML~\citep{plappert2016kit} & \redcheck & \redcheck & \greencheck & \greencheck & 12 & 0.8M & 4K & 71 & 99.0 & 99.6 & 1,623   \\
HumanML3D~\citep{guo2022humanml3d}  & \redcheck & \redcheck & \greencheck & \greencheck  & 29 & 2M & 14K & 147 & 140.0 & 57.50 & 5,371   \\
Motion-X++~\citep{zhang2025motion} & \redcheck & \redcheck & \greencheck & \greencheck & 181 & 19.5M & 120K & 152 & 167.9 & 125.33 & 8,116   \\
\midrule
E.T.~\citep{courant2024exceptional} & \greencheck & \greencheck & \yellowcheck & \redcheck  & 120 & 11M & 115K & 75 & 93.9 & 73.8 & 1,790  \\
\rowcolor{rowcolor!50}
\dataname (Ours) & \greencheck & \greencheck & \greencheck & \greencheck & 314 & 22M & 193K & 107 & 117.3 & 63.6 & 4,599   \\
\bottomrule
\end{tabular}
}
\end{table}

\begin{table*}[t]
    \centering
    \begin{minipage}[t]{0.48\textwidth}
        \centering
        \caption{\small \textbf{Motion refinement and text-motion alignment.} We report metrics on the \dataname{} dataset, comparing raw extracted motions~\citep{wang2024tram} with our refined motions. Captions are generated either from human motions using m2t model~\citep{jiang2024motiongpt} or from RGB frames using our VLM-based approach~\citep{bai2025qwen25vl}.}
\label{tab:refinement_etv2}
\centering
\resizebox{\textwidth}{!}{%
    \begin{tabular}{ll @{\hspace{4mm}} r@{\hspace{2mm}}r@{\hspace{2mm}}r@{\hspace{2mm}}r}
        \toprule
        \textbf{Motion} & \textbf{Caption} & \textbf{TMR-Score} $\uparrow$ & \textbf{R1} $\uparrow$   & \textbf{R2} $\uparrow$   & \textbf{R3}   $\uparrow$  \\
        \midrule
        Extracted       & M2T              & 4.08    & 1.16 & 2.47 & 3.63  \\
        Extracted       & VLM              & 8.06    & 3.65 & 6.64 & 9.20  \\
        \midrule
        Refined         & M2T              & 8.54    & 2.29 & 4.24 & 5.77  \\
        \rowcolor{rowcolor!50}
        Refined         & VLM              & \textbf{16.22}   & \textbf{4.84} & \textbf{8.86} & \textbf{12.34} \\
        \bottomrule
    \end{tabular}}

    \end{minipage}%
    \hfill
    \begin{minipage}[t]{0.48\textwidth}
        \centering
        \caption{\small \textbf{Motion refinement and motion quality.} We compare \dataname{} motion samples with HumanML3D~\citep{guo2022humanml3d}, evaluating raw extracted motions~\citep{wang2024tram} against our refined motions, using either m2t captions from human motions~\citep{jiang2024motiongpt} or our VLM-based captions from RGB frames~\citep{bai2025qwen25vl}.}
\label{tab:refinement_hml3d}
\centering
\resizebox{\textwidth}{!}{%
    \begin{tabular}{lc @{\hspace{4mm}} c@{\hspace{2mm}}c@{\hspace{2mm}}c@{\hspace{2mm}}c@{\hspace{2mm}}c}
        \toprule
        \textbf{Motion} & \textbf{Caption} & $\text{\textbf{FD}}_{\text{TMR}}$ $\downarrow$ & \textbf{P} $\uparrow$ & \textbf{R} $\uparrow$ & \textbf{D} $\uparrow$ & \textbf{C} $\uparrow$ \\
        \midrule
        Extracted       & -                & 595.39                 & 0.53    & 0.13 & 0.32  & 0.15   \\
        Refined         & M2T              & 505.45                 & 0.50    & 0.19 & 0.30  & 0.17   \\
        \rowcolor{rowcolor!50}
        Refined         & VLM              & \textbf{447.69}                 & \textbf{0.55}    & \textbf{0.21} & \textbf{0.37}  & \textbf{0.21}   \\
        \bottomrule
    \end{tabular}}

    \end{minipage}
\end{table*}

Training a joint human–camera model requires paired data of human motions and camera trajectories. 
However, as shown in Table~\ref{tab:datasets}, most prior works provide only one modality, focusing either on human (\emph{e.g.}, HumanML3D~\citep{guo2022humanml3d}) or on camera (\emph{e.g.}, RealEstate10k~\citep{zhou2018realestate10k}). 
More recently, the E.T. dataset~\citep{courant2024exceptional} provides paired samples, but it prioritizes camera aspect, with lower-quality human motions and missing rich textual captions, making it inappropriate to train human motion models.
This motivates us to introduce the \dataname{} dataset, a joint human-camera dataset with good-quality human motions along with motion captions. 
\newline We give an overview of \dataname{} in Section~\ref{sec:data-description} and detail the extraction pipeline in Section~\ref{sec:data-extraction}.

\subsection{Dataset description and comparison}
\label{sec:data-description}

Table~\ref{tab:datasets} compares \dataname{} with existing human and camera datasets. 
Our dataset stands out by providing all modalities, camera trajectories and captions, human motions and captions, while most prior datasets cover only a subset. 
With 193K samples and 314 hours, \dataname{} is also the largest, nearly doubling the number of samples in E.T. (115K) and surpassing other motion-centric datasets such as Motion-X++ (120K). 
In terms of temporal coverage, \dataname{} exhibits longer sequences, with a median length of 107 frames, a mean of 117.3 frames, and a standard deviation of 63.6, indicating both richer and more diverse motion content compared to previous datasets.

\subsection{Extraction pipeline}
\label{sec:data-extraction}

\textbf{Human-camera pair extraction.} Following the E.T. extraction pipeline, we use TRAM~\citep{wang2024tram} to obtain 3D human-camera poses from videos of the CondensedMovies dataset~\citep{bain2020cmd}, and apply the same post-processing steps (filtering, smoothing and cropping to a max-
imum length of 300 frames.
For \dataname{}, we replace SLAHMR~\citep{ye2023slahmr} with TRAM because it is significantly faster ($\sim$1 fps vs.\ $<$0.1 fps), enabling large-scale processing. Moreover, TRAM provides higher-quality estimates, allowing us to keep trajectories that the E.T. pipeline previously filtered.
\newline \textbf{Human-camera captions generation.}
We generate detailed human motion captions inspired by HumanML3D~\citep{guo2022humanml3d} using the Qwen2.5-VL~\citep{bai2025qwen25vl} vision-language model, prompted with video clips and bounding boxes of the target character. The VLM follows annotation guidelines similar to HumanML3D.
To assess the captioning quality we compute the motion-text alignment metrics (\emph{i.e.} cosine similarity and retrieval recall) based on the TMR features~\citep{petrovich2023tmr}. As shown in Table~\ref{tab:refinement_etv2}, our method achieves higher text-motion alignment than existing motion-to-text models~\citep{jiang2024motiongpt}, attaining a TMR-Score of 8.06 against 4.08.
\newline For camera captions, we follow the E.T. methodology: performing motion tagging and inputting it to a large language model (LLM) to produce user-friendly descriptions.
\newline \textbf{Human motion refinement.}
Outputs of TRAM often contain lower-quality human motion compared to mocap-based datasets like HumanML3D. To address this, we introduce a refinement step to enhance motion quality.
The main source of error in TRAM arises from partial observations; e.g., close-up shots capturing only the upper body.
Therefore, to improve overall motion quality, we detect out-of-frame body parts via camera reprojection and refine these regions using the RePaint editing method~\citep{lugmayr2022repaint} with a HumanML3D-pretrained diffusion model.
To assess the captioning quality we compute the motion quality metrics (\emph{i.e.} Fréchet distance and PRDC~\citep{ferjad2020prdc}) based on the TMR features~\citep{petrovich2023tmr}. As shown in Table~\ref{tab:refinement_hml3d}, this step significantly reduces the FD\textsubscript{TMR} score from 595.39 to 447.69. 
\newline We provide additional details on the dataset extraction pipeline in the Appendix~\ref{sec:detail-data}.

\section{Experiments}
\label{sec:experiments}
\subsection{Experimental Setup}
\label{sec:xp-setup}

\textbf{Data representation.} 
{\text{ }}
\newline
\textbf{Framing features ($\rvz_{\text{raw}}$).} We use the 2D Normalized Device Coordinates (NDC): screen-projected coordinates normalized to the range $[-1, 1]$, of nine key human joints (ankles, pelvis, spine, head, shoulders, and wrists). For a sequence of $F$ frames, this yields $\mX_\text{framing} \in \mathbb{R}^{F\times18}$.
\newline 
\textbf{Human features ($\rvx_{\text{raw}}$).} We use the features introduced in~\cite{petrovich2024stmc}, for a motion of $F$ frames: $\mX_\text{human} = (\vr_z, \dot{\vr}_x, \dot{\vr}_y, \dot{\boldsymbol{\alpha}}, \boldsymbol{\Theta}, \mJ) \in \mathbb{R}^{F\times199}$ where $\vr_z \in \mathbb{R}^{F}$ is the Z (up) coordinate of the pelvis, $\dot{\vr}_x \in \mathbb{R}^{F}$ and $\dot{\vr}_y \in \mathbb{R}^{F}$ are the linear velocities of the pelvis, $\dot{\boldsymbol{\alpha}} \in \mathbb{R}^{F}$ is the angular velocity of the Z angle of the body, $\boldsymbol{\Theta} \in \mathbb{R}^{F\times132}$ are the $22$ first SMPL~\citep{loper2023smpl} pose parameters (6D representation~\citep{zhou2019continuity}), and $\mJ \in \mathbb{R}^{F\times63}$ are the $22$ joints positions (pelvis excluded).
\newline
\textbf{Camera feature ($\rvy_{\text{raw}}$).} We extend the features from~\cite{courant2024exceptional} by notably adding the intrinsics. For a trajectory of $F$ frames: $\mX_\text{cam} = (\mR, \dot{\mT}, \mD, \mF) \in \mathbb{R}^{F\times14}$ where $\mR \in \mathbb{R}^{F\times6}$ denotes rotation using the 6D continuous representation~\citep{zhou2019continuity}, $\dot{\mT} \in \mathbb{R}^{F\times3}$ is the linear velocity, $\mD \in \mathbb{R}^{F\times3}$ is the relative distance bewteen the camera and the human, and $\mF \in \mathbb{R}^{F\times2}$ encodes the horizontal and vertical fields of view (assuming the principal point lies at the image center).

\textbf{Metrics.}
{\text{ }}
\newline 
\textbf{Framing metrics.} Since no existing metrics assess framing quality, we propose two metrics based on the 9-joint NDC representation introduced above. 
First, the Fréchet distance $\text{FD}_{\text{framing}}$ measures how well the on-screen framing of the generated camera and human matches a reference distribution. 
Second, the \textit{Out-rate} is the fraction of frames where none of the 9 joints appear on-screen.
\newline
\textbf{Human metrics.} We use the standard text-to-motion metrics~\citep{guo2020action2motion} using the TMR~\citep{petrovich2023tmr} feature space. We then report FD\textsubscript{TMR} and TMR-Score, and TMR-based R-precision. In addition, to evaluate how well generated samples span the variety of real data we compute the TMR-based coverage~\citep{ferjad2020prdc}. 
\newline 
\textbf{Camera metrics.} We use the metrics introduced in~\cite{courant2024exceptional}. To evaluate the camera trajectory quality, we report the FD\textsubscript{CLaTr} and CLaTr-based coverage~\citep{ferjad2020prdc}; to evaluate the camera trajectory coherence, we report the CLaTr-Score and segmentation F1. 

\subsection{Comparison to the state of the art}

\begin{table}[t]
    \caption{\small \textbf{State-of-the-art comparison on the mixed subset.} We compare five baselines: human-conditioned camera generation \director{}~\cite{courant2024exceptional}, independent modality generation \indxy{}, dual-modality generation \jxy{}, triplet-modality generation \jxyz{}, and ReDi~\citep{kouzelis2025boosting}, along with our auxiliary sampling (\ours{}). Results are reported for DiT~\citep{peebles2023scalable} and MAR~\citep{li2024mar}. Superscript $\pm$ denotes the 95\% confidence interval over 10 samplings.}
    \label{tab:sota-mixed}
    \centering
    \resizebox{\textwidth}{!}{%
    \begin{tabular}{l @{\hspace{4mm}} rrrr @{\hspace{4mm}} rrrr @{\hspace{4mm}} rr}
    \toprule
    \multirow{2}{*}{\centering \textbf{Methods}} & \multicolumn{2}{c}{\textbf{Framing}} & \multicolumn{4}{c}{\textbf{Human}} & \multicolumn{4}{c}{\textbf{Camera}} \\
    \cmidrule(r{4mm}){2-3} \cmidrule(r{4mm}){4-7} \cmidrule{8-11}
    & \multicolumn{1}{c}{$\text{FD}_{\text{framing}}\downarrow$}
    & \multicolumn{1}{c}{Out-rate $\downarrow$}
    & \multicolumn{1}{c}{$\text{FD}_{\text{TMR}}\downarrow$}
    & \multicolumn{1}{l}{TMR-Score $\uparrow$}
    & \multicolumn{1}{c}{R3 $\uparrow$}
    & \multicolumn{1}{c}{Coverage $\uparrow$}
    & \multicolumn{1}{c}{$\text{FD}_{\text{CLaTr}}\downarrow$}
    & \multicolumn{1}{c}{CLaTr-Score $\uparrow$}
    & \multicolumn{1}{c}{F1 $\uparrow$}
    & \multicolumn{1}{c}{Coverage $\uparrow$} \\
    \midrule
    Ground-truth & 0.00$^{\textcolor{white}{\pm 0.00}}$ & 0.89$^{\textcolor{white}{\pm 0.00}}$ & 0.00$^{\textcolor{white}{\pm 0.00}}$ & 17.72$^{\textcolor{white}{\pm 0.00}}$ & 22.00$^{\textcolor{white}{\pm 0.00}}$ & 1.00$^{\textcolor{white}{\pm 0.00}}$ & 0.00$^{\textcolor{white}{\pm 0.00}}$ & 68.88$^{\textcolor{white}{\pm 0.00}}$ & 87.43$^{\textcolor{white}{\pm 0.00}}$ & 1.00$^{\textcolor{white}{\pm 0.00}}$ \\
    Auto-encoder & 0.23$^{\textcolor{white}{\pm 0.00}}$ & 4.61$^{\textcolor{white}{\pm 0.00}}$ & 124.78$^{\textcolor{white}{\pm 0.00}}$ & 18.16$^{\textcolor{white}{\pm 0.00}}$ & 21.81$^{\textcolor{white}{\pm 0.00}}$ & 85.30$^{\textcolor{white}{\pm 0.00}}$ & 15.64$^{\textcolor{white}{\pm 0.00}}$ & 57.98$^{\textcolor{white}{\pm 0.00}}$ & 67.04$^{\textcolor{white}{\pm 0.00}}$ & 86.64$^{\textcolor{white}{\pm 0.00}}$ \\
    \midrule
    \multicolumn{10}{l}{DiT} \\
    \quad \director & 22.21$^{\pm 0.03}$ & 60.56$^{\pm 0.14}$ & - & - & - & - & 95.46$^{\pm 0.82}$ & 24.44$^{\pm 0.15}$ & 27.00$^{\pm 0.14}$ & 50.50$^{\pm 0.29}$ \\
    \quad \indxy{} & 11.21$^{\pm 0.12}$ & 48.02$^{\pm 0.24}$ & 357.99$^{\pm 0.52}$ & 25.03$^{\pm 0.06}$ & 4.34$^{\pm 0.12}$ & 10.55$^{\pm 0.17}$ & 67.76$^{\pm 0.20}$ & 46.74$^{\pm 0.11}$ & 46.71$^{\pm 0.24}$ & 53.66$^{\pm 0.36}$  \\
    \rowcolor{rowcolor!25}\quad \ourindxy{} (ours) & 8.24$^{\pm 0.07}$ & 41.24$^{\pm 0.24}$ & 422.45$^{\pm 0.78}$ & 26.46$^{\pm 0.07}$ & 4.64$^{\pm 0.10}$ & 9.08$^{\pm 0.15}$ & 56.41$^{\pm 0.32}$ & 50.69$^{\pm 0.10}$ & 50.72$^{\pm 0.14}$ & 51.39$^{\pm 0.22}$  \\
    \quad \jxy{} & 4.90$^{\pm 0.05}$ & 25.98$^{\pm 0.24}$ & 372.61$^{\pm 0.90}$ & 23.50$^{\pm 0.07}$ & 3.67$^{\pm 0.08}$ & 10.72$^{\pm 0.15}$ & 87.07$^{\pm 0.87}$ & 30.75$^{\pm 0.17}$ & 34.28$^{\pm 0.27}$ & 51.62$^{\pm 0.40}$  \\
    \quad \jxyz{} & 4.18$^{\pm 0.03}$ & 23.88$^{\pm 0.19}$  & 390.08$^{\pm 1.20}$ & 23.88$^{\pm 0.12}$ & 3.22$^{\pm 0.11}$ & 11.58$^{\pm 0.13}$ & 97.45$^{\pm 0.61}$ & 23.34$^{\pm 0.16}$ & 27.40$^{\pm 0.18}$ & 50.80$^{\pm 0.44}$  \\
    \rowcolor{rowcolor!25}\quad \jxyz{}+\ours{} (ours) & 3.76$^{\pm 0.03}$ & 13.90$^{\pm 0.22}$ & 532.42$^{\pm 1.00}$ & 24.58$^{\pm 0.05}$ & 6.13$^{\pm 0.12}$ & 6.88$^{\pm 0.18}$ & 106.97$^{\pm 1.03}$ & 24.61$^{\pm 0.20}$ & 27.43$^{\pm 0.21}$ & 43.36$^{\pm 0.34}$ \\
    \quad ReDi & 5.57$^{\pm 0.04}$ & 28.99$^{\pm 0.22}$ & 360.07$^{\pm 1.26}$ & 22.48$^{\pm 0.06}$ & 5.68$^{\pm 0.18}$ & 12.83$^{\pm 0.16}$ & 83.66$^{\pm 1.05}$ & 22.73$^{\pm 0.22}$ & 26.53$^{\pm 0.20}$ & 55.24$^{\pm 0.41}$  \\
    \rowcolor{rowcolor!50}\quad \ourxy{} (ours) & 3.37$^{\pm 0.02}$ & 16.76$^{\pm 0.19}$ & 431.54$^{\pm 1.15}$ & 25.05$^{\pm 0.07}$ & 3.89$^{\pm 0.14}$ & 8.91$^{\pm 0.13}$ & 80.08$^{\pm 0.76}$ & 32.81$^{\pm 0.19}$ & 36.06$^{\pm 0.25}$ & 48.68$^{\pm 0.20}$  \\
    \midrule
    \multicolumn{10}{l}{MAR} \\
    \quad \indxy{} & 11.59$^{\pm 0.08}$ & 51.05$^{\pm 0.24}$ & 296.01$^{\pm 0.73}$ & 21.71$^{\pm 0.09}$ & 11.69$^{\pm 0.12}$ & 17.48$^{\pm 0.23}$ & 111.42$^{\pm 0.75}$ & 51.96$^{\pm 0.11}$ & 51.69$^{\pm 0.11}$ & 49.85$^{\pm 0.41}$  \\
    \rowcolor{rowcolor!25}\quad \ourindxy{} (ours) & 9.13$^{\pm 0.07}$ & 47.30$^{\pm 0.22}$ & 308.90$^{\pm 0.65}$ & 24.12$^{\pm 0.08}$ & 12.07$^{\pm 0.16}$ & 14.64$^{\pm 0.19}$ & 91.85$^{\pm 0.69}$ & 55.75$^{\pm 0.10}$ & 54.78$^{\pm 0.18}$ & 48.67$^{\pm 0.21}$  \\
    \quad \jxy{} & 8.51$^{\pm 0.07}$ & 40.75$^{\pm 0.28}$ & 275.30$^{\pm 0.55}$ & 21.68$^{\pm 0.06}$ & 10.60$^{\pm 0.19}$ & 17.10$^{\pm 0.28}$ & 117.77$^{\pm 0.63}$ & 42.84$^{\pm 0.14}$ & 42.69$^{\pm 0.23}$ & 54.89$^{\pm 0.37}$  \\
    \quad \jxyz{} & 8.66$^{\pm 0.09}$ & 37.50$^{\pm 0.17}$ & 268.41$^{\pm 0.71}$ & 20.13$^{\pm 0.08}$ & 10.59$^{\pm 0.11}$ & 19.83$^{\pm 0.33}$ & 148.12$^{\pm 0.96}$ & 38.58$^{\pm 0.10}$ & 38.34$^{\pm 0.09}$ & 51.74$^{\pm 0.41}$  \\
    \rowcolor{rowcolor!25}\quad \jxyz{}+\ours{} (ours) & 6.48$^{\pm 0.07}$ & 30.19$^{\pm 0.26}$ & 288.23$^{\pm 0.84}$ & 22.71$^{\pm 0.08}$ & 11.27$^{\pm 0.11}$ & 16.26$^{\pm 0.28}$ & 143.10$^{\pm 1.01}$ & 41.03$^{\pm 0.16}$ & 40.71$^{\pm 0.21}$ & 49.68$^{\pm 0.35}$ \\
    \quad ReDi & 6.96$^{\pm 0.07}$ & 32.25$^{\pm 0.18}$ & 275.58$^{\pm 0.66}$ & 20.84$^{\pm 0.08}$ & 10.90$^{\pm 0.11}$ & 18.41$^{\pm 0.32}$ & 122.40$^{\pm 0.77}$ & 42.60$^{\pm 0.15}$ & 42.70$^{\pm 0.21}$ & 54.96$^{\pm 0.51}$  \\
    \rowcolor{rowcolor!50}\quad \ourxy{} (ours) & 6.42$^{\pm 0.04}$ & 33.65$^{\pm 0.23}$ & 301.39$^{\pm 0.25}$ & 24.46$^{\pm 0.07}$ & 11.28$^{\pm 0.09}$ & 14.14$^{\pm 0.14}$ & 108.74$^{\pm 0.46}$ & 45.96$^{\pm 0.14}$ & 45.39$^{\pm 0.22}$ & 53.67$^{\pm 0.38}$  \\
    \bottomrule
    \end{tabular}}
\end{table}
\begin{figure}[t]
\centering
\begin{minipage}{.49\textwidth}
  \centering
  \captionsetup[subfigure]{skip=0.0pt}
  \subcaptionbox{\small $\text{FD}_{\text{framing}}$-TMR-Score\label{fig:quant-mixed-dit-clatr}}[.49\linewidth]{
    \input{fig/mixed-comparisons/mixed-dit-tmr}
  }
  \subcaptionbox{\small $\text{FD}_{\text{framing}}$-CLaTr-Score\label{fig:quant-mixed-dit-tmr}}[.49\linewidth]{
    \input{fig/mixed-comparisons/mixed-dit-clatr}
    }
  \caption{\small \textbf{Comparison in DiT on the mixed set.} Framing quality and modality-text alignment for $\vc$ guidance ranges from 5 to 11. The optimal region is at the bottom-right (low framing FD, high alignment).}
  \label{fig:quant-mixed-dit}
\end{minipage}%
\hfill
\begin{minipage}{.49\textwidth}
  \centering
  \captionsetup[subfigure]{skip=0.0pt}
  \subcaptionbox{\small $\text{FD}_{\text{framing}}$-TMR-Score\label{fig:quant-mixed-mar-clatr}}[.49\linewidth]{
    \input{fig/mixed-comparisons/mixed-mar-tmr}
  }
  \subcaptionbox{\small $\text{FD}_{\text{framing}}$-CLaTr-Score\label{fig:quant-mixed-mar-tmr}}[.49\linewidth]{
    \input{fig/mixed-comparisons/mixed-mar-clatr}
  }

  \caption{\small \textbf{Comparison in MAR on the mixed set.}  Framing quality and modality-text alignment for $\vc$ guidance ranges from 1 to 5. The optimal region is at the bottom-right (low framing FD, high alignment).}
  \label{fig:quant-mixed-mar}
\end{minipage}
\end{figure}

In this section, we compare our \textbf{auxiliary sampling} (\ours{}) method against five baselines: (1) \textbf{human-conditioned camera generation} \director{}~\cite{courant2024exceptional}: a two-stage approach in which one model generates human motions, and a second model generates camera trajectories conditioned on those motions; (2) \textbf{independent modality generation} \indxy{}: two separate models for each modality; (3) \textbf{dual-modality generation} \jxy{}: a single model generates both modalities without the auxiliary modality; (4) \textbf{triplet-modality generation}  \jxyz{}: a single model generates both modalities and the auxiliary modality; and (5) \textbf{ReDi}~\citep{kouzelis2025boosting}: a single model for both modalities and the auxiliary modality, with representation sampling leveraging the auxiliary modality. Except for \director{} (originally trained with DiT), we evaluate all baselines and our method using both DiT-based~\citep{peebles2023scalable} and MAR~\citep{li2024mar} architectures (see Appendix~\ref{sec:supp-xp} for more details on architectures).

\textbf{Quantitative results.} 
Table~\ref{tab:sota-mixed} reports a comparison of our auxiliary sampling (\ours{}) method against state-of-the-art baselines across both DiT and MAR architectures on the mixed subset. 
We summarise our experimental observations as follows:
\newline \textbf{(i) Auxiliary sampling consistently improves framing (multimodal coherence).}
Applying \ours{} leads to systematic improvements in framing quality ($\text{FD}_{\text{framing}}$) and out-of-frame (Out-rate) rates across all baseline configurations and architectures.
For DiT, auxiliary sampling reduces $\text{FD}_{\text{framing}}$ from $11.21\!\to\!8.24$ for \indxy{}, from $4.90\!\to\!3.37$ when applied to \jxy{}. A similar trend holds for MAR, where $\text{FD}_{\text{framing}}$ decreases from $11.59\!\to\!9.13$ for \indxy{} and from $8.51\!\to\!6.42$ for \jxy{}.
Out rates follow the same pattern: for DiT, \ours{} reduces the out rate from $48.02\%\!\to\!41.24\%$ (\indxy{}) and from $25.98\%\!\to\!16.76\%$ (\jxy{}), while for MAR it decreases from $51.05\%\!\to\!47.30\%$ and $40.75\%\!\to\!33.65\%$, respectively. 
Overall, auxiliary sampling achieves the best $\text{FD}_{\text{framing}}$ and out rates across both architectures (DiT and MAR): showing that it is architecture-agnostic and consistently enhances framing quality, i.e. multimodal coherence, for all settings (independent and joint).
\newline \textbf{(ii) Conditioning on human motion alone is insufficient for strong framing.}
We next examine the human-conditioned camera generation baseline \director{}, where human trajectories are first generated using the same backbone as the independent setting (\indxy{}), and camera motion is subsequently conditioned on these synthesized humans. 
While this strategy is weaker than other baselines: for example, under DiT, \director{} yields a $\text{FD}_{\text{framing}}$ of $22.21$, compared to $8.24$ for \indxy{}+\ours{} and $3.37$ for \jxy{}+\ours{}. Its out rate is also higher with $60.56\%$ against $41.24\%$ and $16.76\%$, respectively. 
These results that conditioning on generated human motion alone provides limited contextual information for accurate camera framing. 
In contrast, joint generation methods, especially when combined with \ours{}, consistently outperform \director{} and the independent baseline (\indxy{}), highlighting the importance of jointly modeling modalities during sampling.
\newline \textbf{(iii) Auxiliary sampling strengthens per-modality alignment while preserving quality.}
Beyond framing, \ours{} improves text–modality alignment across both human and camera dimensions. 
For human motion, auxiliary sampling increases TMR-Score from $23.50\!\to\!25.05$ (DiT, \jxy{}) and from $21.68\!\to\!24.46$ (MAR, \jxy{}), while also improving R3.
For camera motion, CLaTr-Score improves from $30.75\!\to\!32.81$ (DiT, \jxy{}) and from $42.84\!\to\!45.96$ (MAR, \jxy{}), along with gains in F1. 
Although modality generation quality metrics ($\text{FD}_{\text{TMR}}$ and $\text{FD}_{\text{CLaTr}}$) slightly increase, these changes are modest and do not outweigh the substantial gains in alignment and framing consistency. 
Overall, \ours{} achieves a favorable trade-off, improving multimodal coherence and per-modality alignment while maintaining strong generation quality across architectures.
\begin{figure}[t]
\centering
\begin{minipage}{.49\textwidth}
  \centering
  \begin{overpic}[width=1.0\linewidth]{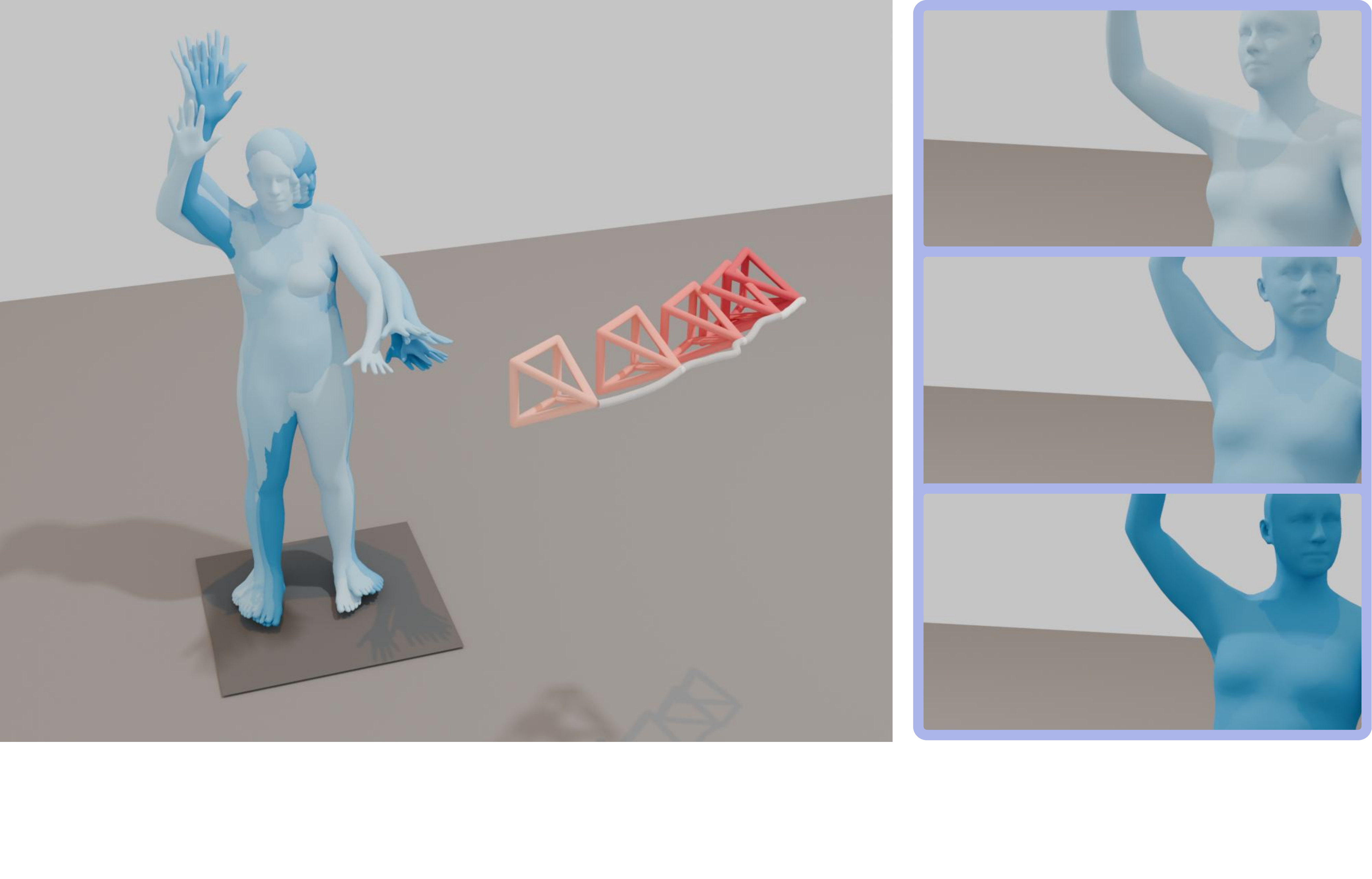}
    \put(0,5){\small \textbf{Human:} A person raising their right arm.}
    \put(0,0){\small \textbf{Camera:} The camera performs a trucking right.}
  \end{overpic}
  \captionof{figure}{\small \textbf{Example with DiT on the mixed subset.}}
  \label{fig:qual-mixed-dit}
\end{minipage}%
\hfill
\begin{minipage}{.49\textwidth}
  \centering
  \begin{overpic}[width=1.0\linewidth]{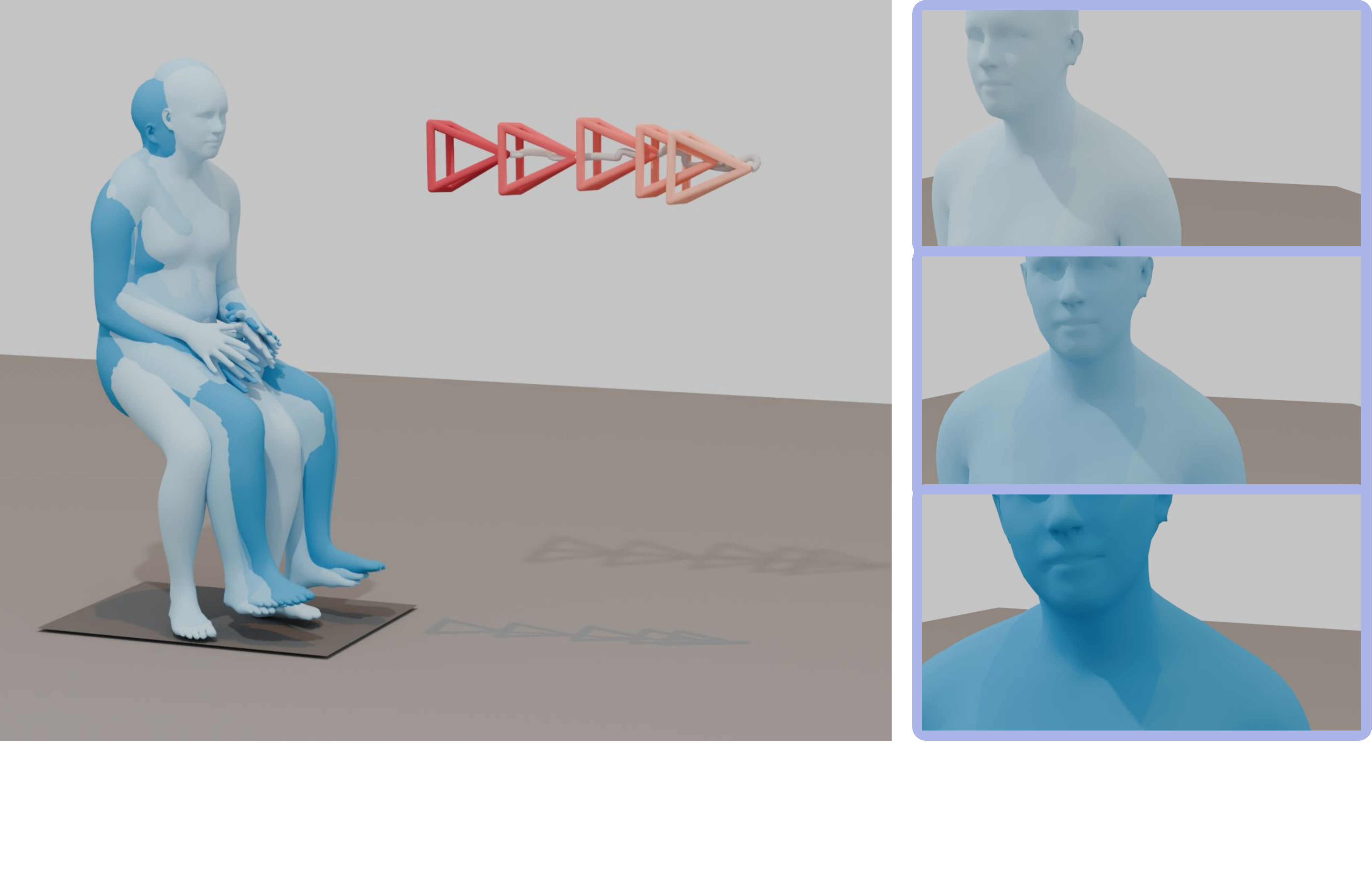}
    \put(0,5){\small \textbf{Human:} A person sitting.}
    \put(0,0){\small \textbf{Camera:} The camera performs a push in.}
  \end{overpic}
  \captionof{figure}{\small \textbf{Example with MAR on the mixed subset.} 
  }
  \label{fig:qual-mixed-mar}
\end{minipage}
\end{figure}
\newline Moreover, we compare our method with baselines for DiT and MAR in Figures~\ref{fig:quant-mixed-dit} and~\ref{fig:quant-mixed-mar}, showing the trade-off between framing quality ($\text{FD}_{\text{framing}}$) and modality-text alignment (TMR for human, CLaTr for camera) across different textual guidance values ($w_c$ in Equation~\ref{eq:epsilon-cfg-z}). The optimal point lies in the bottom-right corner of each plot (low $\text{FD}_{\text{framing}}$, high modality scores). 
Across both architectures and both modalities, our auxiliary sampling achieves the best performance, improving both framing quality and textual alignment, showing its effectiveness and generality.

\textbf{Qualitative results.} 
Figures~\ref{fig:qual-mixed-dit} and~\ref{fig:qual-mixed-mar} show qualitative results with ours sampling for DiT and MAR, respectively. 
In both cases, the generated human motion is precise, for example, for DiT, the person raises the \emph{right} arm as specified. 
The camera trajectories are also coherent with the prompt, and accurately following the human motion while maintaining correct on-screen framing, with the subject’s head consistently in view. 
These results highlight that \ours{} produces humans and cameras that are well aligned with the input prompts, achieving precise motion and coherent framing across different architectures.
Further examples are provided on our \href{https://www.lix.polytechnique.fr/vista/projects/2025_pulpmotion_courant/}{project page} and in the Appendix~\ref{sec:supp-gen}.

\subsection{Ablation study} 

\begin{table}[htp]
    \caption{\small \textbf{Auxiliary guidance ablation on the mixed subset.} We vary the auxiliary guidance weight $w_z$ to evaluate its effect on the framing, camera and human metrics. Results are reported for DiT~\citep{peebles2023scalable} and MAR~\citep{li2024mar}. Superscript $\pm$ denotes the 95\% confidence interval over 10 samplings.} 
    
    \label{tab:ablation-mixed}
    \centering
    \resizebox{\textwidth}{!}{%
    \begin{tabular}{c @{\hspace{4mm}} rrrr @{\hspace{4mm}} rrrr @{\hspace{4mm}} rr}
    \toprule
    \multirow{2}{*}{\centering \textbf{$\bm{w_z}$}} & \multicolumn{2}{c}{\textbf{Framing}} & \multicolumn{4}{c}{\textbf{Human}} & \multicolumn{4}{c}{\textbf{Camera}} \\
    \cmidrule(r{4mm}){2-3} \cmidrule(r{4mm}){4-7} \cmidrule{8-11}
    & \multicolumn{1}{c}{$\text{FD}_{\text{framing}}\downarrow$}
    & \multicolumn{1}{c}{Out-rate $\downarrow$}
    & \multicolumn{1}{c}{$\text{FD}_{\text{TMR}}\downarrow$}
    & \multicolumn{1}{l}{TMR-Score $\uparrow$}
    & \multicolumn{1}{c}{R3 $\uparrow$}
    & \multicolumn{1}{c}{Coverage $\uparrow$}
    & \multicolumn{1}{c}{$\text{FD}_{\text{CLaTr}}\downarrow$}
    & \multicolumn{1}{c}{CLaTr-Score $\uparrow$}
    & \multicolumn{1}{c}{F1 $\uparrow$}
    & \multicolumn{1}{c}{Coverage $\uparrow$} \\
    \midrule
    \multicolumn{10}{l}{DiT} \\
    \quad 0.00 & 4.90$^{\pm 0.05}$ & 25.98$^{\pm 0.24}$ & 372.61$^{\pm 0.90}$ & 23.50$^{\pm 0.07}$ & 3.67$^{\pm 0.08}$ & 10.72$^{\pm 0.15}$ & 87.07$^{\pm 0.87}$ & 30.75$^{\pm 0.17}$ & 34.28$^{\pm 0.27}$ & 51.62$^{\pm 0.40}$ \\
    \rowcolor{rowcolor!50}\quad 0.25 & 3.37$^{\pm 0.02}$ & 16.76$^{\pm 0.19}$ & 431.54$^{\pm 1.15}$ & 25.05$^{\pm 0.07}$ & 3.89$^{\pm 0.14}$ & 8.91$^{\pm 0.13}$ & 80.08$^{\pm 0.76}$ & 32.81$^{\pm 0.19}$ & 36.06$^{\pm 0.25}$ & 48.68$^{\pm 0.20}$ \\
    \quad 0.50 & 3.09$^{\pm 0.02}$ & 11.99$^{\pm 0.16}$ & 493.53$^{\pm 1.64}$ & 25.30$^{\pm 0.07}$ & 7.23$^{\pm 0.10}$ & 7.09$^{\pm 0.17}$ & 90.06$^{\pm 0.58}$ & 32.45$^{\pm 0.13}$ & 35.98$^{\pm 0.09}$ & 44.98$^{\pm 0.31}$ \\
    \quad 0.75 & 3.37$^{\pm 0.02}$ & 9.66$^{\pm 0.12}$ & 548.60$^{\pm 1.62}$ & 24.99$^{\pm 0.07}$ & 7.08$^{\pm 0.09}$ & 5.63$^{\pm 0.15}$ & 123.16$^{\pm 0.75}$ & 29.58$^{\pm 0.12}$ & 30.88$^{\pm 0.17}$ & 38.88$^{\pm 0.34}$ \\
    \midrule
    \multicolumn{10}{l}{MAR} \\
    \quad 0.00 & 8.51$^{\pm 0.07}$ & 40.75$^{\pm 0.28}$ & 275.30$^{\pm 0.55}$ & 21.68$^{\pm 0.06}$ & 10.60$^{\pm 0.19}$ & 17.10$^{\pm 0.28}$ & 117.77$^{\pm 0.63}$ & 42.84$^{\pm 0.14}$ & 42.69$^{\pm 0.23}$ & 54.89$^{\pm 0.37}$ \\
    \rowcolor{rowcolor!50}\quad 0.50 & 6.42$^{\pm 0.04}$ & 33.65$^{\pm 0.23}$ & 301.39$^{\pm 0.25}$ & 24.46$^{\pm 0.07}$ & 11.28$^{\pm 0.09}$ & 14.14$^{\pm 0.14}$ & 108.74$^{\pm 0.46}$ & 45.96$^{\pm 0.14}$ & 45.39$^{\pm 0.22}$ & 53.67$^{\pm 0.38}$ \\
    \quad 1.00 & 5.93$^{\pm 0.02}$ & 32.09$^{\pm 0.17}$ & 326.29$^{\pm 0.31}$ & 25.42$^{\pm 0.07}$ & 11.35$^{\pm 0.14}$ & 12.57$^{\pm 0.12}$ & 144.02$^{\pm 0.60}$ & 44.05$^{\pm 0.16}$ & 40.19$^{\pm 0.15}$ & 47.26$^{\pm 0.34}$ \\
    \quad 1.50 & 6.04$^{\pm 0.02}$ & 32.77$^{\pm 0.15}$ & 346.14$^{\pm 0.42}$ & 25.65$^{\pm 0.05}$ & 11.35$^{\pm 0.20}$ & 11.63$^{\pm 0.19}$ & 193.14$^{\pm 0.46}$ & 40.61$^{\pm 0.11}$ & 36.64$^{\pm 0.17}$ & 38.21$^{\pm 0.41}$ \\
    \bottomrule
\end{tabular}}
\end{table}
\vspace{-0.4cm}
\begin{figure}[htp]
\centering
\begin{minipage}{.49\textwidth}
  \centering
  \captionsetup[subfigure]{skip=0.0pt}
  \subcaptionbox{\small $\text{FD}_{\text{framing}}$-TMR-Score\label{fig:ablation-mixed-dit-tmr}}[.49\linewidth]{
    \input{fig/mixed-ablations/mixed-dit-tmr}
  }
  \subcaptionbox{\small $\text{FD}_{\text{framing}}$-CLaTr-Score\label{fig:ablation-mixed-dit-clatr}}[.49\linewidth]{
    \input{fig/mixed-ablations/mixed-dit-clatr}
  }
  \caption{\small \textbf{$w_z$ ablation in DiT on the mixed set.} Framing quality and modality-text alignment for $\vc$ guidance ranges from 4 to 12. The optimal region is at the bottom-right (low framing FD, high alignment).}
  \label{fig:ablation-mixed-dit}
\end{minipage}%
\hfill
\begin{minipage}{.49\textwidth}
  \centering
  \captionsetup[subfigure]{skip=0.0pt}
  \subcaptionbox{\small $\text{FD}_{\text{framing}}$-TMR-Score\label{fig:ablation-mixed-mar-tmr}}[.49\linewidth]{
    \input{fig/mixed-ablations/mixed-mar-tmr}
  }
  \hfill
  \subcaptionbox{\small $\text{FD}_{\text{framing}}$-CLaTr-Score\label{fig:ablation-mixed-mar-clatr}}[.49\linewidth]{
    \input{fig/mixed-ablations/mixed-mar-clatr}
  }

  \caption{\small \textbf{$w_z$ ablation in MAR on the mixed set.} Framing quality and modality-text alignment for $\vc$ guidance ranges from 1 to 5. The optimal region is at the bottom-right (low framing FD, high alignment).}
  \label{fig:ablation-mixed-mar}
\end{minipage}
\end{figure}

To assess controllability and effectiveness of auxiliary sampling, we ablate in \Cref{tab:ablation-mixed} the auxiliary guidance weight $w_z$ (Eq~(\ref{eq:epsilon-cfg-z})) on both DiT and MAR.
We see that a \textbf{(1) moderate guidance weight improves framing and text–modality alignment}. On DiT, increasing $w_z$ from $0.00$ to $0.25$ reduces $\text{FD}_{\text{framing}}$ $4.90\!\to\!3.37$ and Out-rate $25.98\!\to\!16.76$; on MAR, $w_z{=}0.50$ lowers them $8.51\!\to\!6.42$ and $40.75\!\to\!33.65$. \textbf{(2) Pushing $w_z$ further keeps improving framing but degrades fidelity}: $\text{FD}_{\text{TMR}}$ and $\text{FD}_{\text{CLaTr}}$ rise (DiT $431.54\!\to\!493.53$, MAR $301.39\!\to\!326.29$). \textbf{(3) At high weights, it becomes unstable} ($w_z{=}0.75$ DiT, $1.50$ MAR), with $\text{FD}_{\text{TMR}}$ spiking to $548.60$ and $\text{FD}_{\text{CLaTr}}$ to $193.14$.
\newline We then illustrate Figures~\ref{fig:ablation-mixed-dit} and~\ref{fig:ablation-mixed-mar} for the trade-off between framing quality ($\text{FD}_{\text{framing}}$, lower is better) and text–modality alignment (TMR, CLaTr; higher is better) as the \ours{} guidance weight $w_z$ varies. The optimum lies near the bottom-right of each plot. Across both architectures, we see: (1) introducing guidance yields a large gain: $w_z{:}0\!\to\!0.25$ (DiT) and $0.50$ (MAR) shift points toward the bottom-right; (2) further increases, $0.50$ (DiT), $1.0$ (MAR), continue to improve framing but begin to reduce fidelity, reflected by larger markers (higher Fréchet distances); and (3) at very high weights, $0.75$ (DiT), $1.50$ (MAR), performance degrades on both axes.

\textbf{Summary of findings.} From our experiments and ablations, we find that: (1) our proposed sampling consistently improves human–camera coherence (better framing, fewer empty frames) while preserving strong per-modality performance; and 
(2) the gains generalise across architectures, though absolute performance depends on tuning the guidance weight (as with CFG).

\section{Conclusion}
\label{sec:conclusion}
In this paper, we presented a unified framework for joint generation of human motion and camera trajectories, enforcing multimodal coherence via an auxiliary modality: the on-screen framing. Extensive evaluations on the proposed \dataname{} dataset demonstrate the generality and effectiveness of our auxiliary sampling. 
Our approach currently supports character-level framing but does not yet provide fine-grained control over specific body parts or localized regions, which we identify as an important direction for future work. More broadly, our auxiliary sampling approach is general and could be extended to other modalities and application domains.

\section*{Acknowledgements} 
{This work was supported by ANR/France 2030 program (\textit{ANR-22-CE23-0007}, \textit{ANR-22-CE39-0016}, \textit{ANR-23-IACL-0005}), Hi!Paris grant, fellowship and chair, DATAIA Convergence Institute as part of the “Programme d’Investissement d’Avenir” (ANR-17-CONV-0003) operated by Ecole Polytechnique, IP Paris,  Inria Action Exploratoire PREMEDIT (Precision Medicine using Topology), and the CIEDS. 
\newline This project was granted access to the IDRIS High-Performance Computing (HPC) resources under the allocation 2024-AD011013951R2 made by GENCI. 
\newline We sincerely thank Yuanzhi Zhu and Nicolas Dufour, for their insightful discussions that contributed to this work and their meticulous proofreading.}

\newpage
\bibliography{short-strings,references}
\bibliographystyle{iclr2026_conference}

\newpage
\appendix

\startcontents[sections] 
\printcontents[sections]{l}{1}{\setcounter{tocdepth}{3}} 

\section{Use of Large Language Models}
\label{supp:llm_usage}
We used large language models solely for text polishing and grammar correction during manuscript preparation. No LLMs were involved in the conception or design of the method, experiments, or analysis. All technical content, results, and conclusions have been independently and carefully verified and validated by the authors.

\section{Detailed related work}
\label{supp:related work}
\paragraph{Detailed Human motion generation Related Work.}
Inspired by the success of denoising diffusion models in image generation~\citep{ho2020ddpm,rombach2022ldm}, several pioneering works~\citep{tevet2022mdm,kim2023flame,zhang2024motiondiffuse} adapt diffusion processes to human motion generation. These models are then followed by extensions that leverage pre-trained latent spaces for efficiency, apply consistency distillation for faster sampling, improve caption–motion alignment, and exploit external databases for higher-quality motion~\citep{chen2023mdl,dai2024motionlcm,andreou2025lead,zhang2023remodiffuse}.

While diffusion-based approaches typically represent motion data as raw joint positions and orientations, or continuous latent vectors, another line of work adopts Vector Quantization (VQ) with discrete motion tokens. TM2T~\citep{guo2022tm2t} first introduce VQ into text-to-motion generation, followed by T2M-GPT~\citep{zhang2023generating}, which employed a GPT-style autoregressive model~\citep{brown2020gpt}. More recently, MMM~\citep{pinyoanuntapong2024mmm}, MoMask~\citep{guo2024momask}, and BiPO~\citep{hong2024bipo} propose to apply bidirectional attention-based masked generation, inspired by MaskGIT~\citep{chang2022maskgit}.

Recently, the masked autoregressive architecture (MAR)~\citep{li2024mar} has been proposed to combine the strengths of autoregressive and diffusion models. 
It has drawn significant attention in the human motion community~\citep{li2024mar,meng2024mardm,xiao2025motionstreamer}, as it leverages an autoregressive transformer to handle temporal dynamics while retaining the high generation quality of diffusion models, enabling new state-of-the-art performances.

Nevertheless, most motion generation methods treat motion as an isolated modality, which oversimplifies real-world scenarios where humans continuously interact with their surroundings. Consequently, recent research has begun modeling human interactions with objects~\citep{xu2023interdiff,peng2025hoi,geng2025auto}, other humans~\citep{liang2024intergen,fan2024freemotion,shan2024towards}, and scenes~\citep{wang2024move,cen2024generating,jiang2024autonomous}.
However, while recent studies have considered human–camera interaction in motion estimation~\citep{Priyanka20253dv,ye2023slahmr,wang2024tram,kocabas2024pace,Sun_2023_CVPR}, motion generation remains largely unexplored, with existing efforts treating camera parameters merely as constraints or conditioning signals rather than modeling their joint distribution with motion.

\paragraph{Detailed Generative Camera Trajectory generation Related Work.}

Over the past two decades, camera control and generation have progressed from handcrafted, rule-based geometric design~\citep{blinn1988looking,lino2015intuitive,drucker1992cinema} to deep learning methods that exploit the descriptive and fitting capacity of neural networks: approaches that either learn cinematic rules from example-based references~\citep{jiang20sig,jiang21siga} or leverage the differentiability of deep models to optimize camera trajectories in real-data-supported 3D environments~\citep{wang2023jaws,jiang2024cinematic,chen2024dreamcinema}.

However, these example-based methods often rely on carefully curated reference videos, and in some cases even synthetic annotation pairs, either to train discriminative models or to optimize trajectories. To mitigate this dependency, other works explore reinforcement learning (RL). In drone cinematography~\citep{Huang_2019_CVPR,bonatti2020autonomous}, RL is guided by human pose and optical flow, while in indoor environments, \cite{Xie_2023_ICCV} propose to use an aesthetic model as the reward function. Though effective within specific environments, both example-based and RL-based methods often collapse into limited trajectory styles and require environment-specific training, resulting in poor generalization.

Yet, with the rapid progress of image and video generative models~\citep{polyak2024movie,opensora2,wang2025wan}, a notable direction is to bypass explicit 3D representations and instead treat the model as a universal renderer. This has enabled direct camera-controlled video generation~\citep{wang2024motionctrl,he2024cameractrl,xu2024camco,bahmani2024vd3d,zheng2024cami2v,cheong2024boosting,wang2024akira}. While showing great potential, this line of work faces several limitations: (1) it overlooks scene semantics (e.g., character performance); (2) it still relies on manually designed, complex camera trajectories, which remain challenging for users; and (3) given the relatively low quality of current video generators, the outputs are hard to use directly in production, while the end-to-end nature of these models prevents artists from accessing intermediate assets (e.g., meshes, trajectories, lighting conditions).

To achieve geometric controllability and provide intermediate assets without requiring expert-designed trajectories, \citet{jiang2024ccd} introduced the first diffusion-based approach for camera generation. Although limited to synthetic data, their key idea of deriving camera behavior from semantic prompts opens a new direction. Subsequently,~\cite{courant2024exceptional} proposed E.T., a large-scale dataset of realistic camera trajectories with human motion from real films, together with evaluation metrics and novel architectural designs. 
DanceCamAnimator~\cite{wang2024dancecamanimator} and DanceCamera3D~\cite{wang2024dancecamera3d} also focus on dance-specific camera control conditioned on music.
More recently, GenDoP~\citep{zhang2025gendop} constructed an object-wise, interaction-centric dataset and employed an autoregressive model to generate trajectories from textual descriptions and visual inputs. 

Similarly to human motion generation, most camera generation works condition on human motion but rely solely on global trajectories, which restricts the interaction between camera and subject and overlooks the intrinsic joint distribution problem.
In this work, we aim to bridge this gap by addind human motion into the camera trajectory generation pipeline, modelling the symbiosis between how and what to film.

\section{Detailed method}
\label{supp:method}
\def\dd{\,\mathrm{d}}
\def\im{\mathrm{im}\,}
\def\coim{\mathrm{coim}\,}
\def\ker{\mathrm{ker}\,}
\subsection{Background}
\subsubsection{Moore-Penrose Pseudo-Inverse and induced projections}\label{sec:pseudo_inverse}
We provide the background defining the Moore-Penrose pseudo-inverse of an $m\times n$ matrix $W$. Let $k := \min\{n, m\}$.

Consider a singular value decomposition of $W$, given by $W = UDV^\top$, where:
\begin{enumerate}
	\item $U$ is an $m\times m$ orthogonal matrix, i.e., $U^\top U = \mI_m$,
	\item $V$ is an $n\times n$ orthogonal matrix, i.e., $V^\top V = \mI_n$,
	\item $D = \mathrm{diag}(d_{1}, \ldots, d_{k})$ is a $k\times k$ diagonal matrix with non-negative, non-increasing diagonal entries.
\end{enumerate}

The Moore-Penrose pseudo-inverse $W^\dagger$ is then given by:
\[
	W^\dagger = V D^\dagger U^\top,
\]
where $D^\dagger$ is the $k\times k$ diagonal matrix with entries $D^\dagger_{i,i} = d_{i}^{-1}$ if $d_{i} > 0$, and $0$ otherwise.

Note that if the $k\times k$ matrix $W^\top W$ is invertible, then $D^\dagger = D^{-1}$ and hence $W^\dagger = W^\top(W^\top W)^{-1}$.

Define $P := D^\dagger D$, a diagonal matrix with entries:
\[
	P_{i,i} = \begin{cases}
		1 & \text{if } d_{i} \neq 0, \\
		0 & \text{otherwise}.
	\end{cases}
\]
We end up with the following properties:
\begin{enumerate}
	\item \textit{Projection:} $W^\dagger W = V P V^\top$, and $(W^\dagger W)^2 = W^\dagger W$.
	\item \textit{Symmetry:} $(W^\dagger W)^\top = W^\dagger W$.
	\item \textit{Projection image:} $\ker W = \mathrm{im}(\mI_n - W^\dagger W)$.
\end{enumerate}
Thus, $P_\perp := \mI_n - W^\dagger W$ is the orthogonal projection onto $\ker
W$, and $P_{\newparallel} := W^\dagger W$ is the orthogonal projection onto
the orthogonal of $\ker W$: 
the coimage of $W$.
\subsubsection{Statistics}
We start by a well-known special case of the main theorem from
\cite{cochranDistributionQuadraticForms1934}, that we apply to projection matrices.
\begin{theorem}[Cochran]\label{thm:cochran}
	Let $X \sim \mathcal{N}(\mu, \sigma^2 \mI_n)$ be an isotropic Gaussian random vector, and let $F \subseteq \mathbb{R}^n$ be a linear subspace.
	If $P_F$ and $P_{F^\perp}$ are the respective orthogonal projections onto $F$ and its orthogonal complement $F^\perp$, then:
	\begin{enumerate}
		\item \label{enum:cochran:law}$P_F X \sim \mathcal{N}(P_F \mu, \sigma^2 P_F)$ and $P_{F^\perp} X \sim \mathcal{N}(P_{F^\perp} \mu, \sigma^2 P_{F^\perp})$ are (possibly degenerate) Gaussian random vectors.
		\item \label{enum:cochran:indep}$P_F X$ and $P_{F^\perp} X$ are independent.
	\end{enumerate}
\end{theorem}
For completeness, we provide a succinct proof below.
\begin{proof}
	Let $P = \begin{bmatrix} P_F \\ P_{F^\perp} \end{bmatrix}$. 
		By the properties of Gaussian vectors, $PX$ is a Gaussian vector with covariance matrix:
\begin{equation}\label{eq:cochran_sigma}
	\Sigma =
	\sigma^2
	\begin{bmatrix}
		P_F\circ P_F^\top          & P_F\circ P_{F^\top}^\top        \\
		P_{F^\top}\circ P_{F}^\top & P_{F^\top}\circ P_{F^\top}^\top
	\end{bmatrix}
	=
	\sigma^2
	\begin{bmatrix}
		P_F & 0          \\
		0   & P_{F^\top}
	\end{bmatrix},
\end{equation}
	where the last inequality follows from the properties of the orthogonal
	projections $P_F$ and $P_{F^\perp}$, namely: $P_F = P_F^\top$, $P_F^2 = P_F$,
	and $P_F P_{F^\perp} = 0$.
	Similarly, since the multiplication by $P$ is linear, $\mathbb{E}(PX) = \begin{bmatrix} P_F \mu \\
	P_{F^\perp} \mu \end{bmatrix}$. \Cref{enum:cochran:law} follows from the block decomposition,
	and \Cref{enum:cochran:indep} follows from \Cref{eq:cochran_sigma} and the fact that
	uncorrelated Gaussian vectors are independent.
\end{proof}

\subsection{Method}\label{sec:theory:method}
Let $\mW^\dagger$ denote the Moore-Penrose pseudo-inverse of $\mW$ (see
\Cref{sec:pseudo_inverse}).
Using this notation, the matrix $\mP_\perp := \mI - \mW^\dagger
\mW$ (resp. $\mP_{\newparallel} := \mW^\dagger\mW$) can be then identified as the orthogonal
projection onto $\ker \mW$ (resp. $\ker (\mW)^\perp$).
As $\rvu = [\rvx, \rvy]^\top \sim \mathcal N(\mm, \sigma^2 \mI_n)$, Cochran's theorem (\Cref{thm:cochran}) then
guarantees that the corresponding projections $\rvu$:
\begin{equation}\label{eq:p-distributions2}
              \rvu_\perp := \mP_\perp \rvu \sim \mathcal N(\mP_\perp \mm, \sigma^2 \mP_\perp)
              \quad \textnormal{and} \quad
              \rvu_{\newparallel} := \mP_{\newparallel} \rvu \sim \mathcal N(\mP_{\newparallel} \mm, \sigma^2 \mP_{\newparallel}),
          \end{equation}
are independent (possibly degenerate) Gaussian vectors.

Observe that $\rvz = \mW\rvu =
\mW\rvu_{\newparallel}$, so $\rvz$ is
$\rvu_{\newparallel}$-measurable and thus independent of
$\rvu_\perp$.
Moreover, $\rvu_{\newparallel} = \mW^\dagger
\mW\rvu = \mW^\dagger \rvz$, which shows that
$\rvu_{\newparallel}$ is a measurable function of $\rvz$.
Therefore,
\begin{equation}\label{eq:Eu|z}
    \mathbb E[\rvu \mid \rvz] 
		= \mathbb E\left[\rvu_\perp + \mW^\dagger \rvz \mid \rvz\right] 
		\overset{\rvu_\perp \textnormal{ indep. } \rvz}{=} 
		\mathbb E(\rvu_\perp) + \mathbb E[\mW^\dagger \rvz \mid \rvz] \overset{\rvz\textnormal{-meas.}}{=}
		\mP_\perp \mm + \mW^\dagger \rvz.
\end{equation}
We have the following decomposition of $\rvu$ into two independent variables:
$\rvu = \mP_{\perp}\rvu + \mP_{\newparallel}\rvu  = \mP_{\perp}\rvu + \mW^\dagger \rvz$.
This induces a decomposition of the density $p_\rvu$ of $\rvu$ into two density
functions\footnotemark{} $p_{\rvu_\perp}$ and
$p_{\rvu_{\parallel}}=p_{\mW^\dagger \rvz}$
of $\rvu_\perp$ and $\rvu_{\newparallel} = \mW^\dagger \rvz$ respectively. 
\footnotetext{%
    Note that $\rvu_\perp$ and $\rvu_{\newparallel}$ do not admit densities
    w.r.t. the Lebesgue measure on $\mathbb{R}^n$,
    since they are supported on the non-full-dimensional vector spaces $\ker \mW$ and $\ker (\mW)^\perp$ respectively.
    Nevertheless, they admit densities $p_{\rvu_\perp}$ and $p_{\rvu_{\newparallel}}$ w.r.t.
    the respective Lebesgue measures on these subspaces. See \Cref{sec:bayes_density} for more details.
}
Given point $u\in \mathbb{R}^n$:
\begin{equation}\label{eq:decomposition}
	p_\rvu(u) 
	= p_{(\rvu_{\perp},\rvu_{\parallel})}(u_\perp,u_{\newparallel})
	\overset{\textnormal{indep.}}= p_{\rvu_\perp}(u_\perp)p_{\rvu_{\newparallel}}(u_{\newparallel}),
\end{equation}
where the functions $p_{\rvu_\perp}$ and $p_{\rvu_{\newparallel}}$ are given by:
\begin{align*}
	v\mapsto p_{\rvu_\perp}(v) =&\frac {\ind{\ker \mW}(v)} {\sqrt{2\pi\sigma^2}^{\dim \ker \mW}}  \exp\left[  -\frac {1} {2\sigma^2} (v-\mm)^\top (v-\mm) \right],\textnormal{ and}\\
	v\mapsto p_{\rvu_{\newparallel}}(v) =&
	\frac {\ind{\ker (\mW)^\perp}(v)} {\sqrt{2\pi\sigma^2}^{\dim \ker (\mW)^\perp}} \exp\left[ -\frac {1} {2\sigma^2} (v-\mm)^\top (v-\mm)  \right].
\end{align*}
\subsection{More details on the density decomposition}\label{sec:bayes_density}
We give here more details on the possible decompositions of the density $p_\rvu$ of  $\rvu$.
The decomposition given in \Cref{eq:decomposition} from  ensures that we have for a given point $u\in \mathbb R^n$:
\begin{equation}\label{eq:decomposition2}
	p_\rvu(u) = p_{\rvu_\perp}( (\mI - \mW^\dagger \mW)u)p_{\rvu_{\newparallel}}(\mW^\dagger \mW u).
\end{equation}

Following the argumentation of \Cref{eq:Eu|z}, given a point
$z\in\mathbb{R}^m$, the conditional distribution $\rvu\mid \rvz=z$ of $\rvu$
conditionally to the event $ \left\{ \rvz=z \right\} $ is given by $\rvu_\perp
+ \mW^\dagger z$: 
 a translation of $\rvu_\perp$ 
by the constant $\mW^\dagger z$
and thus admits a density $p_{\rvu\mid \rvz=z}$,
given by:
\begin{equation*}
	(u,z)\longmapsto p_{\rvu\mid\rvz=z}(u) = p_{\rvu_\perp+\mW^\dagger z}(u) =  p_{\rvu_\perp}(u-\mW^\dagger z).
\end{equation*}
This can be directly shown by the change of variable $u=u_\perp +
\mW^\dagger z = u_\perp + \mW^\dagger \mW u_{\newparallel}$ in \Cref{eq:decomposition2}, 
since on one hand, we have $\rvz=\mW \rvu$ almost surely, hence
\begin{equation*}
	p_{\rvu}(u) 
	= p_{(\rvu_{\perp},\rvu_{\parallel})}(u_\perp,u_{\newparallel})
	\overset{\textnormal{indep.}}= p_{\rvu_{\perp}}(u_\perp)p_{{\rvu_{\newparallel}}}(u_{\newparallel})
	\overset{\textnormal{shift}} = p_{\rvu_\perp + u_{\newparallel}}(u_\perp + u_{\newparallel}) p_{\rvu_{\newparallel}}(u_{\newparallel}).
\end{equation*}
Furthermore, on the other hand, we have $\rvu_{\newparallel} = \mW^\dagger \rvz$ and $\rvz =
\mW \rvu$, hence $p_\rvz(\cdot) \propto p_{\rvu_{\newparallel}}(\mW \cdot)$,
and therefore, for any point $u$ and $z= \mW u$:

\begin{equation*}
	p_{\rvu}(u) 
	\overset{z=\mW u} = p_{\rvu_\perp + \mW^\dagger z}(u)p_{\rvu_\parallel}(\mW^\dagger z)
\overset{z=\mW u} \propto p_{\rvu_\perp + \mW^\dagger z}(u)  p_{\rvz}(z).
\end{equation*}

This equality being true for (almost) every $u$ and $z=\mW u$, we conclude that $p_{\rvu\mid\rvz=z} \propto p_{\rvu_\perp + \mW^\dagger z}$ almost everywhere.
This property can be shown more directly by invoking the fact that the
$\sigma$-algebra generated by $\rvz$ and the one of $\rvu_{\newparallel}$ are
the same.

\subsection{Auxiliary sampling derivation}
\label{sec:aux-derivation}

In this section, we detail the derivation from \Cref{supp:eq:cfg-z} to \Cref{eq:epsilon-cfg-z}.
Starting from \Cref{eq:cfg} and applying the decomposition in \Cref{eq:decomposition-lemma}, we have:
\begin{equation}
    \begin{split}
        \nabla_{\rvx_t, \rvy_t} \log \tilde{p}(\rvx_t, \rvy_t| \rc)
         & = \nabla_{\rvx_t, \rvy_t} \log p(\rvu_{\perp}) 
         + (1 + w_z) \nabla_{\rvx_t, \rvy_t} \log p(\rvu_{\newparallel})                \\
         & \quad +  w_c (\nabla_{\rvx_t, \rvy_t} \log p(\rvx_t, \rvy_t | \rc)
        - \nabla_{\rvx_t, \rvy_t} \log p(\rvx_t, \rvy_t)).
    \end{split}
    \label{supp:eq:cfg-z}
\end{equation}

Therefore, since $\xy \sim \mathcal{N}(\mm, \sigma^2\mI)$ we obtain:
\begin{equation}
    \begin{split}
        \nabla_{\rvx_t, \rvy_t} \log p(\rvx_t, \rvy_t| \rc) 
        &= -\frac{1}{\sigma^2} (\mI - \mP_{\newparallel}) \left(\xyt - \mm\right) -\frac{1}{\sigma^2} (1+w_z) \mP_{\newparallel} \left(\xyt - \mm\right) \\
        &\quad -\frac{1}{\sigma^2}  w_c \left((\rvx_c - \mm_c) - (\xyt - \mm)\right) \\
        &= -\frac{1}{\sigma^2} \left(\xyt - \mm\right) \\
        &\quad -\frac{1}{\sigma^2} w_z \mP_{\newparallel} \left(\xyt - \mm\right) \\
        &\quad -\frac{1}{\sigma^2}  w_c \left((\xytc - \mm_c) - (\xyt - \mm)\right).
    \end{split}
\end{equation}

Finally, recalling that $\ee = - \frac{1}{\sigma}(\rvx - \mm)$, we can express the sampling equation with noise prediction as:
\begin{equation}
    \begin{split}
        \ee \left( \rvx_t, \rvy_t, \rc, t \right)
         & = \ee \left( \rvx_t, \rvy_t, \emptyset, t \right)                                                             \\
         & \quad + w_z \mP_{\newparallel} \; \ee\left(\rvx_t, \rvy_t, \emptyset, t \right)                                          \\
         & \quad + w_c (\ee \left( \rvx_t, \rvy_t, \rc, t \right)  - \ee \left( \rvx_t, \rvy_t, \emptyset, t \right) ).
    \end{split}
\end{equation}

\section{Detailed dataset}
\label{supp:dataset}
\subsection{Detailed pipeline}
\label{sec:detail-data}

\begin{figure}[htbp]
    \centering
    \includegraphics[width=\textwidth]{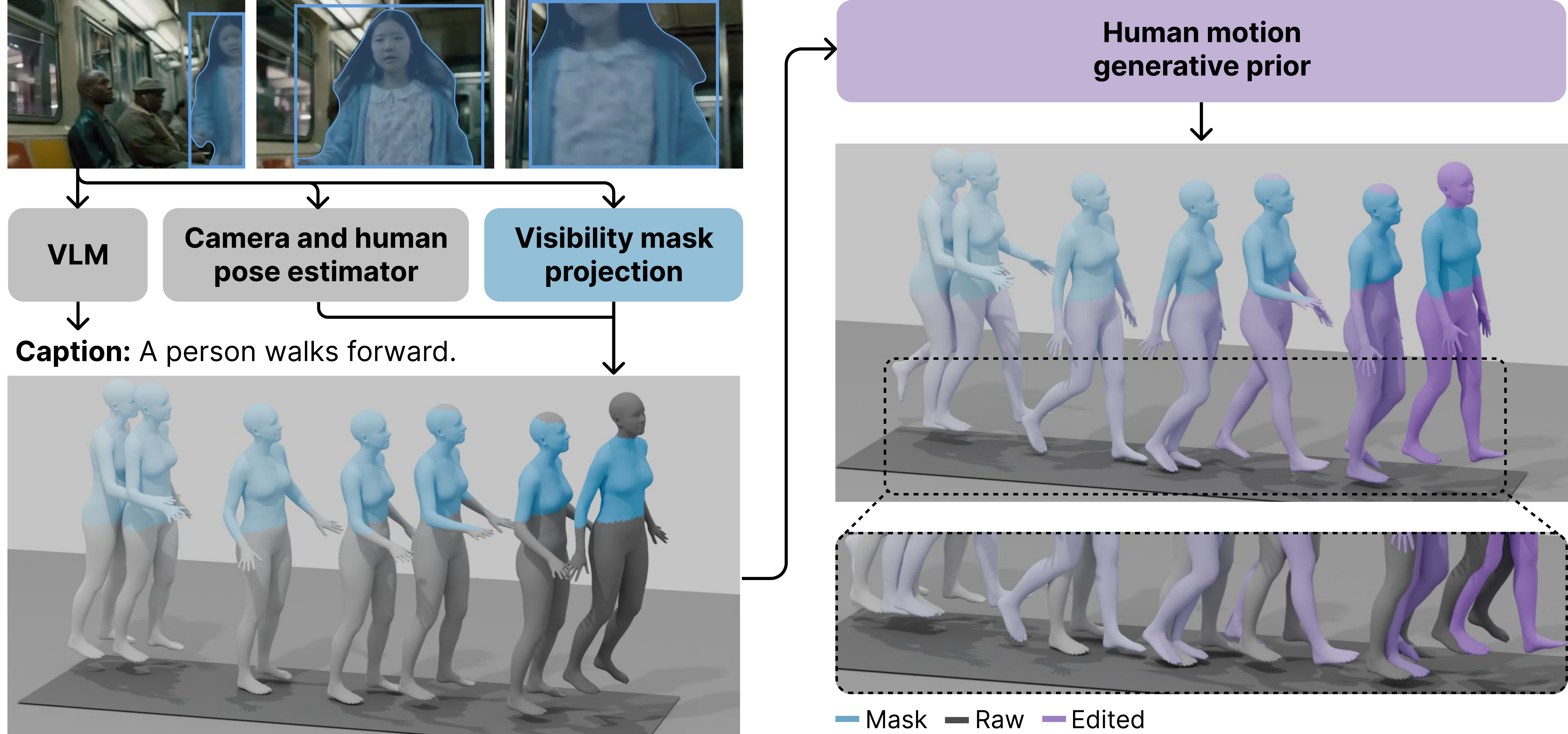}
    \caption{\small \textbf{Dataset refinement pipeline.} Given RGB frames from a video, we first estimate the camera and human pose. We then identify the out-of-screen body parts by reprojection. Finally, we refine the out-of-screen parts using a generative prior.}
    \label{supp:fig:dataset}
\end{figure}

In this section, we describe the construction of the \dataname{} dataset, illustrated in Figure~\ref{supp:fig:dataset}.
We first apply an off-the-shelf camera-human pose estimator~\citep{wang2024tram} to infer both camera and human poses from video clips of the CondensedMovies dataset~\citep{bain2020cmd}. As noted in the main manuscript, a key challenge of video-based pose estimation is handling occluded or unseen body parts, which are often inaccurately predicted.

\begin{figure}[H]
  \centering
\begin{minipage}{\linewidth}
\small
\textbf{Input prompt:}
\begin{figure}[H]
  \centering
  \includegraphics[width=0.9\linewidth]{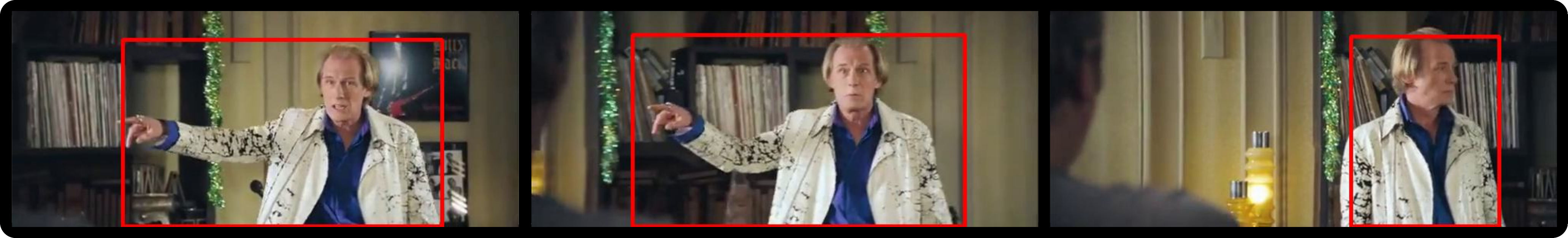}
\end{figure}
\begin{verbatim}
Task:
  Observe the full motion sequence of the highlighted person carefully.
  Summarize their motion in one concise, accurate, and complete sentence 
  in English.
Instructions:
  - Clearly state the primary global actions (e.g., walk, run, jump, 
  sit, squat).
  - Specify relevant details of the motion:
  - Directions (e.g., forward, backward, left, right, clockwise).
  - Body parts involved explicitly (e.g., left arm, right foot, hips).
  - Movement trajectory patterns if present (e.g., circular, zig-zag, 
  straight line).
  - Movement speed or intensity (e.g., slowly, quickly, rapidly).
  - Order and sequencing of actions clearly (what happens first, 
  next, last).
  - Always describe global movement explicitly, even when emphasizing 
  local actions (e.g., "A person stands still and turns their head" 
  instead of just "turns head").
  - Refer only to "a person," avoiding mention of highlighting or boxes.
  Avoid:
  - Inferring emotional states or intentions.
  - Vague or overly generalized descriptions ("A person moves," 
  "A person acts").
  - Excessive spatial precision or measurement details ("moves 
  exactly 5 cm forward").
  Good Examples:
  - "A person rises from sitting, walks quickly in a clockwise circle, 
  then sits down again."
  - "A person jumps repeatedly side-to-side while holding their 
  left wrist with the right hand, then runs rapidly backward."
  - "A person crouches, then climbs upward onto an object 
  positioned to their right."
  Bad Examples (DO NOT PRODUCE):
  - Over-general or no main motion: "A person looks."
  - Too detailed spatially: "A person moves exactly 
  10 centimeters left."
  - Explicit mention of highlighting: "The character in the red box jumps."
  Your response must strictly follow the format of the good examples provided.
  \end{verbatim}
\textbf{Response:}
\begin{verbatim}
A person points forward with their right hand while walking slightly to the left, 
then stops and turns their head to the right.
\end{verbatim}
\end{minipage}  
   \caption{\textbf{Example of human motion caption instruction.}}
  \label{fig:prompt_example}
\end{figure}

To address this, we first identify poorly estimated regions by reprojecting visible joints. We then use a vision-language model (VLM) to generate captions describing human motion, providing bounding boxes around the target person to guide the model’s focus. We show an example of human motion caption generation in \Cref{fig:prompt_example} with input prompt and VLM response.

Next, in the right part of Figure~\ref{supp:fig:dataset}, we refine the occluded regions using a diffusion-based editing method~\citep{lugmayr2022repaint} with a model pretrained on HumanML3D~\citep{guo2022humanml3d}. To avoid artifacts caused by naive editing, we refine the entire sub-kinematic chain of each occluded joint rather than modifying joints in isolation. Since visible parts remain largely unchanged, projection consistency between the reconstructed body and RGB frames is preserved.

\subsection{Detailed statistics}

\begin{table}[t]
    \caption{ 
    \small \textbf{Comparison of \dataname{} and E.T.~\citep{courant2024exceptional} datasets.} We compare the full (\textit{all}), \textit{pure}, and \textit{mixed} subsets of \dataname{} with the E.T.. We summarize modality coverage, available captions, dataset size (hours, frames, samples), sample length statistics (median, mean, std), and vocabulary size.
    }
    \label{tab:subdatasets}
    \centering
    \resizebox{\textwidth}{!}{%
    \begin{tabular}{lrrrrrrrc@{\hspace{1mm}}cc@{\hspace{2mm}}c}
    \toprule
    \multirow{2}{*}{\textbf{Dataset}} & \multicolumn{2}{c}{\textbf{Camera}} & \multicolumn{2}{c}{\textbf{Human}} & \multirow{2}{*}{\textbf{\#Hours}} & \multirow{2}{*}{\textbf{\#Frames}} & \multirow{2}{*}{\textbf{\#Samples}} & \multicolumn{3}{c}{\textbf{Sample lengths (frames)}} & \multirow{2}{*}{\textbf{\#Vocabulary}} \\
    & Traj & Caption & Motion & Caption &        &         &          & Median      & Mean      & Std      &  \\
    \midrule
    \multicolumn{1}{l}{E.T.~\cite{courant2024exceptional}} & \multirow{4}{*}{\greencheck} & \multirow{4}{*}{\greencheck} & \multirow{4}{*}{\yellowcheck} & \multirow{4}{*}{\redcheck}  \\
    \quad  \textit{all} &  &  &  &  & 120 & 11M & 115K & 75 & 93.9 & 73.8 & 1,790  \\
    \quad \textit{pure} &  &  &  &  & 20 & 1.8M & 30K & 46 & 59.5 & 49.1 & 941    \\
    \quad \textit{mixed} &  &  &  &  & 67 & 6M & 65K & 72 & 92.9 & 75.06 & 1,579    \\
    \midrule
    \rowcolor{rowcolor!50}\multicolumn{1}{l}{\dataname{} (Ours)} & \multirow{4}{*}{\greencheck} & \multirow{4}{*}{\greencheck} & \multirow{4}{*}{\greencheck} & \multirow{4}{*}{\greencheck} & \multicolumn{7}{l}{} \\
    \quad  \textit{all} &  &  &  &  & 314 & 22M & 193K & 107 & 117.3 & 63.6 & 4,599  \\
    \quad \textit{pure} &  &  &  &  & 51 & 3.7M & 41K & 70 & 91.1 & 59.2 & 2,831  \\
    \quad \textit{mixed} &  &  &  &  & 170 & 12M & 105K & 108 & 116.44 & 60.44 & 4,143 \\
    \bottomrule
    \end{tabular}}
\end{table}

Table~\ref{tab:subdatasets} compares our \dataname{} dataset with E.T.~\citep{courant2024exceptional} across several dimensions. 

Overall, \dataname{} significantly increases the dataset size, containing 314 hours and 22M frames compared to E.T.’s 120 hours and 11M frames. Our dataset also provides longer samples (median 107 frames vs. 75). 
The “pure” and “mixed” subsets follow the same trends, demonstrating consistent improvements.

Thanks to our refinement pipeline, as shown in the main manuscript, \dataname{} ensures higher-quality human motions.
Additionally, \dataname{} includes HumanML3D-style human motion captions, which are not available in E.T.

\section{Detailed experiments}
\label{supp:experiments}
\subsection{Autoencoder}

\subsubsection{Detailed experimental setup}

\paragraph{Implementation details.}
We adopt the ResNet-based autoencoder from MARDM~\citep{meng2024mardm} with ReLU activations. The joint encoder and three modality-specific decoders have temporal down-/up-sampling by a factor of 4, and each consist of two 1D-ResNet blocks. Latent dimensions are set to $64$ for the camera, $128$ for the human, and $64$ for the projection. The model is trained for 325 epochs on the full Pulp~Motion dataset with $64$-frame samples (and evaluated on $300$-frame samples), using AdamW with a learning rate of $1.9\times10^{-4}$, a batch size of $128$, on a single A100 GPU. A linear warmup of $1$K steps is applied, followed by a decay of 0.1 after $4$K steps.

\subsubsection{Detailed performances}

\begin{table}[H]
    \caption{\small \textbf{Reconstruction evaluation of autoencoder.} We report reconstruction metrics for \textit{pure} and \textit{mixed} subsets. Metrics span projection accuracy (MPJProjE, $\text{FD}_{\text{framing}}$), human pose quality (MPJPE, $\text{FD}{\text{TMR}}$, TMR-Score), and camera alignment (APE, $\text{FD}_{\text{CLaTr}}$, CLaTr-Score).}
    \label{tab:autoencoder}
    \centering
    \resizebox{\textwidth}{!}{%
    \begin{tabular}{l @{\hspace{4mm}} rr @{\hspace{4mm}} rrr @{\hspace{4mm}} rrr}
    \toprule
    \multirow{2}{*}{\centering \textbf{Methods}} & \multicolumn{2}{c}{\textbf{Framing}} & \multicolumn{3}{c}{\textbf{Human}} & \multicolumn{3}{c}{\textbf{Camera}} \\
       \cmidrule(r{4mm}){2-3} \cmidrule(r{4mm}){4-6} \cmidrule{7-9}
       & MPJProjPE $\downarrow$ & $\text{FD}_{\text{framing}}$ $\downarrow$ & MPJPE $\downarrow$ & $\text{FD}_{\text{TMR}}$ $\downarrow$ & TMR-Score $\uparrow$ & APE  $\downarrow$  & $\text{FD}_{\text{CLaTr}}$ $\downarrow$ & CLaTr-Score $\uparrow$ \\
    \hline

    \multicolumn{9}{l}{\textit{pure}} \\
    \quad Ground truth & 0.00 & 0.00 & 0.00 & 0.00 & 16.47 & 0.00 & 0.00 & 70.25 \rule{0pt}{2.6ex}\\
    \rowcolor{rowcolor!50}\quad AE & 0.09 &  0.14 & 3.26 & 105.57 & 15.93 & 0.15 & 19.26 & 60.45 \\
    \midrule
    \multicolumn{9}{l}{\textit{mixed}} \\
    \quad Ground truth & 0.00 & 0.00 & 0.00 & 0.00 & 17.72 & 0.00 & 0.00 & 68.88 \rule{0pt}{2.6ex}\\
    \rowcolor{rowcolor!50}\quad AE &  0.08 & 0.23 & 5.63 & 124.78 & 18.16 & 0.18 & 15.64 & 57.98 \\
    \bottomrule
    \end{tabular}}
\end{table}

We evaluate the reconstruction quality of the autoencoder introduced in~\Cref{sec:latent-space} using modality-specific errors: mean per-joint projected error (MPJProjPE) and mean per-joint error (MPJPE) for human motion, and absolute pose error (APE) for the camera. Additionally, we compute reconstruction Fréchet distances and modality–text alignment metrics from \Cref{sec:xp-setup}.

As shown in Table~\ref{tab:autoencoder}, the autoencoder achieves low MPJProjPE ($0.08$–$0.09$), indicating reliable 2D frame reconstruction across both subsets. 
MPJPE reveals discrepancies in 3D pose recovery, particularly in the mixed setting ($5.63$ vs. $3.26$). 
APE remains low ($\leq 0.18$) but shows slight degradation in the mixed case, consistent with the observed drop in CLaTr-Score.

\subsection{Generation}
\label{sec:supp-gen}

\subsubsection{Detailed experimental setup}
\label{sec:supp-xp}

\begin{figure}[ht]
    \centering
    \hfill
    \begin{minipage}{0.45\textwidth}
        \centering
        \includegraphics[width=\linewidth]{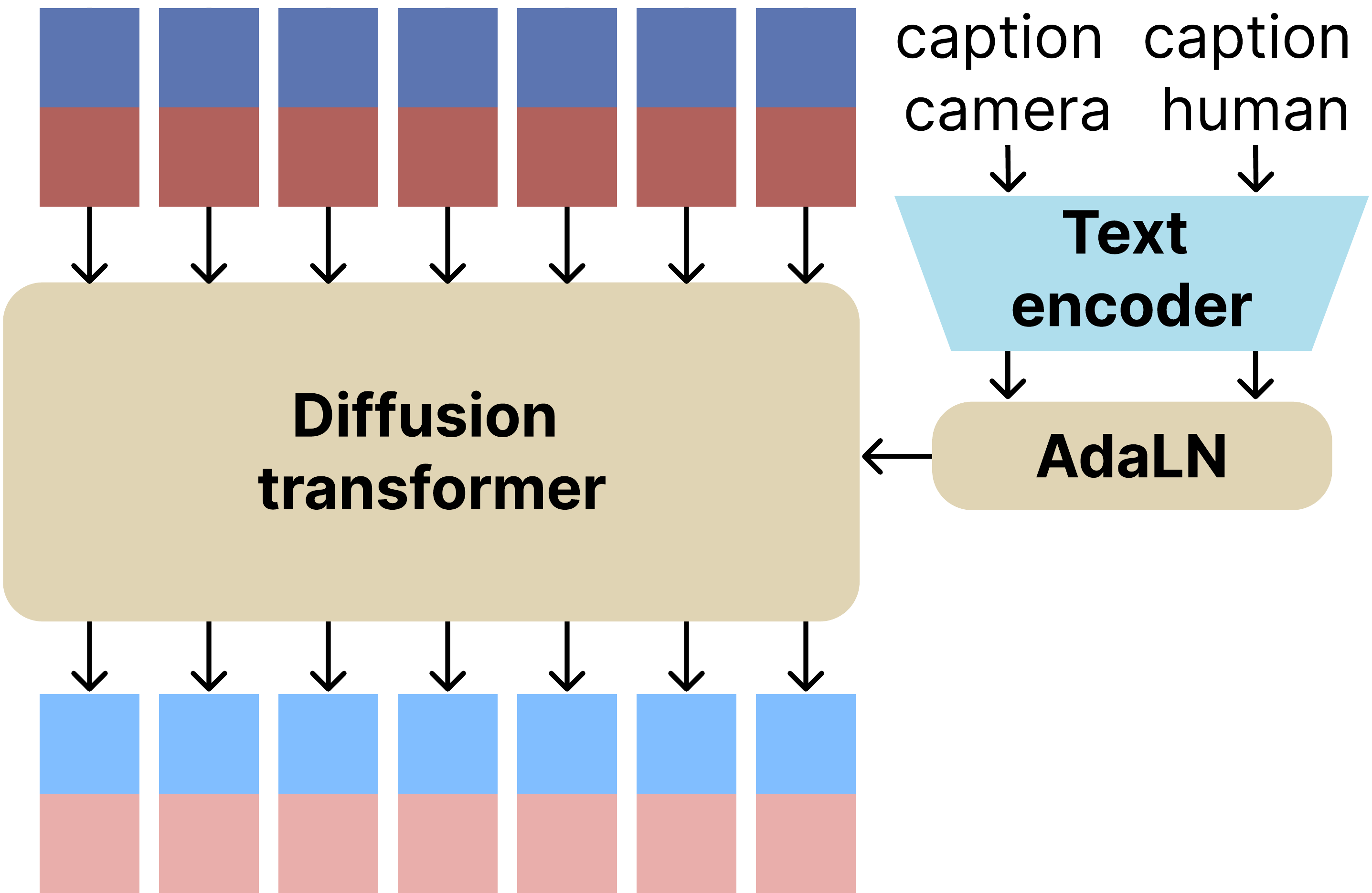}
        \caption{\small \textbf{Overview of the DiT architecture.}}
        \label{fig:mar}
    \end{minipage}\hfill
    \begin{minipage}{0.45\textwidth}
        \centering
        \includegraphics[width=\linewidth]{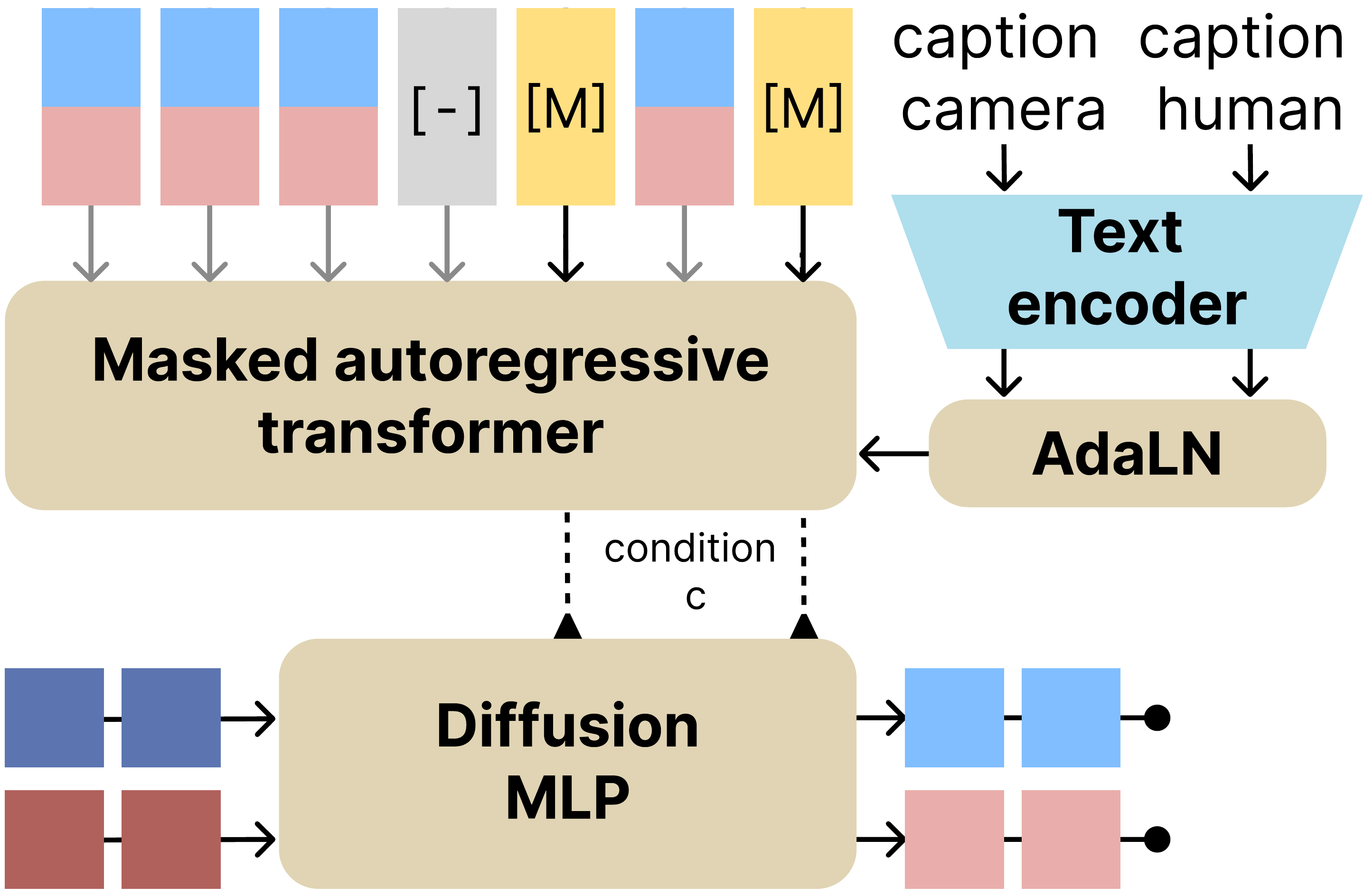}
        \caption{\small \textbf{Overview of the MAR architecture.}}
        \label{fig:dit}
    \end{minipage}\hfill
\end{figure}

We illustrate in Figures~\ref{fig:dit}~and~\ref{fig:mar} both DiT~\cite{peebles2023scalable} and MAR~\cite{li2024mar} architectures used in this work.

\paragraph{Implementation details.}
We evaluate two architectures: a DiT-based model with in-context conditioning~\citep{courant2024exceptional} and a MAR-based model with AdaLN conditioning~\citep{meng2024mardm}. To ensure fairness, both are scaled to $\sim28.3$M parameters. The DiT model has $8$ layers with a hidden dimension $532$ and $14$ attention heads. The MAR model uses a single-layer autoregressive transformer (hidden dimension $512$, $8$ heads) and a diffusion head with $3$ MLP layers of width $1024$. Both models are trained for $93$k steps on the \textit{pure} subset and $330$K steps on the \textit{mixed} subset on the pure and mixed subsets of Pulp~Motion with 300-frame samples, using AdamW with a learning rate of $3\times10^{-4}$, a batch size of $128$, on a single A100 GPU. A linear warmup of $2$K steps is applied, followed by decay by $0.1$ after $50$K steps.
For inference, we perform $50$ DDPM sampling steps.

\subsubsection{More qualitative results on mixed dataset}

\begin{figure}[t]
\centering
\begin{minipage}{.49\textwidth}
  \centering
  \begin{overpic}[width=1.0\linewidth]{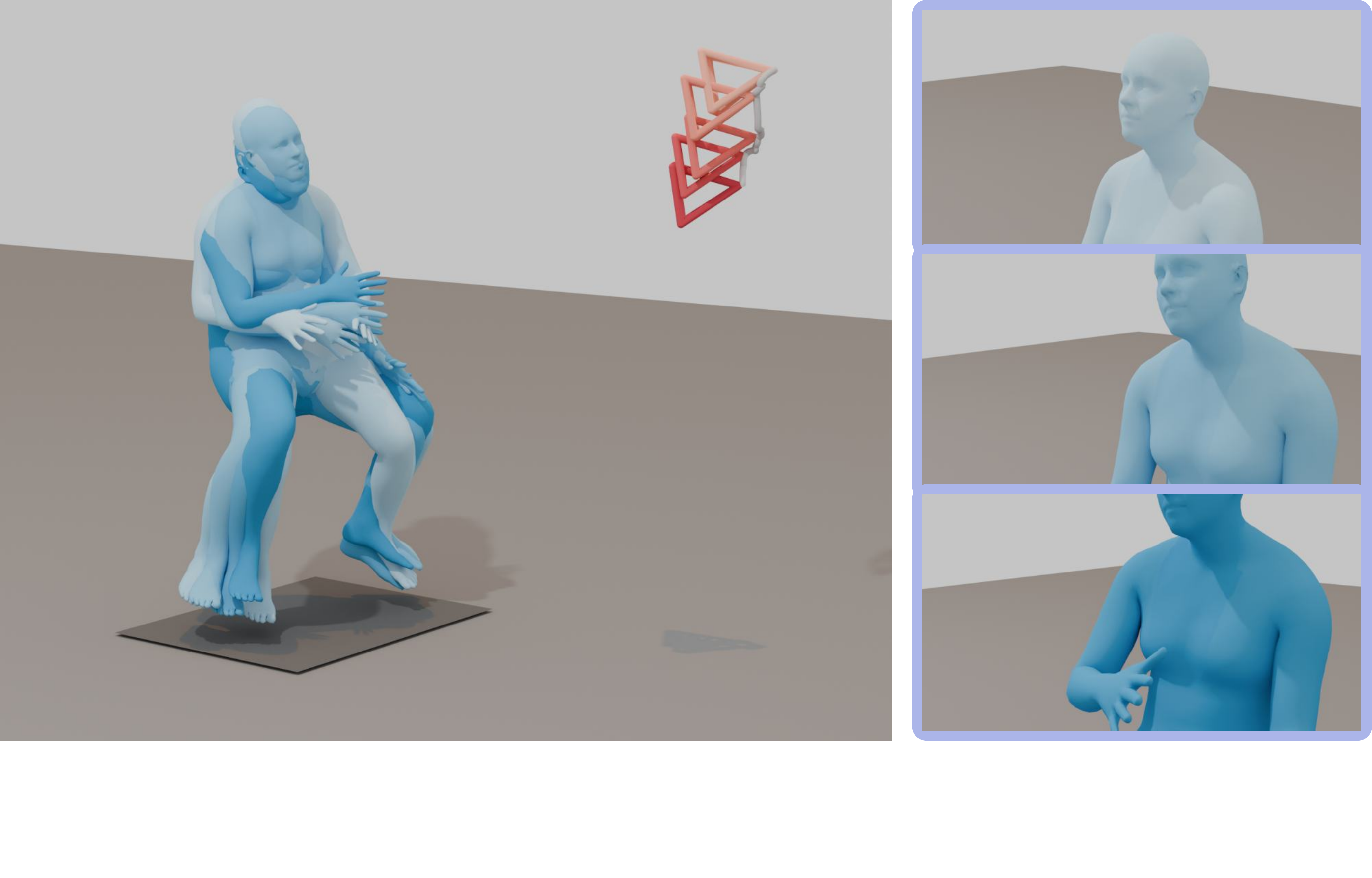}
    \put(0,5){\small \textbf{Human:} A person sitting.}
    \put(0,0){\small \textbf{Camera:} The camera performs a boom bottom.}
  \end{overpic}
  \captionof{figure}{\small \textbf{Example with DiT on the mixed set.}}
  \label{fig:qual-mixed-dit2}
\end{minipage}%
\hfill
\begin{minipage}{.49\textwidth}
  \centering
  \begin{overpic}[width=1.0\linewidth]{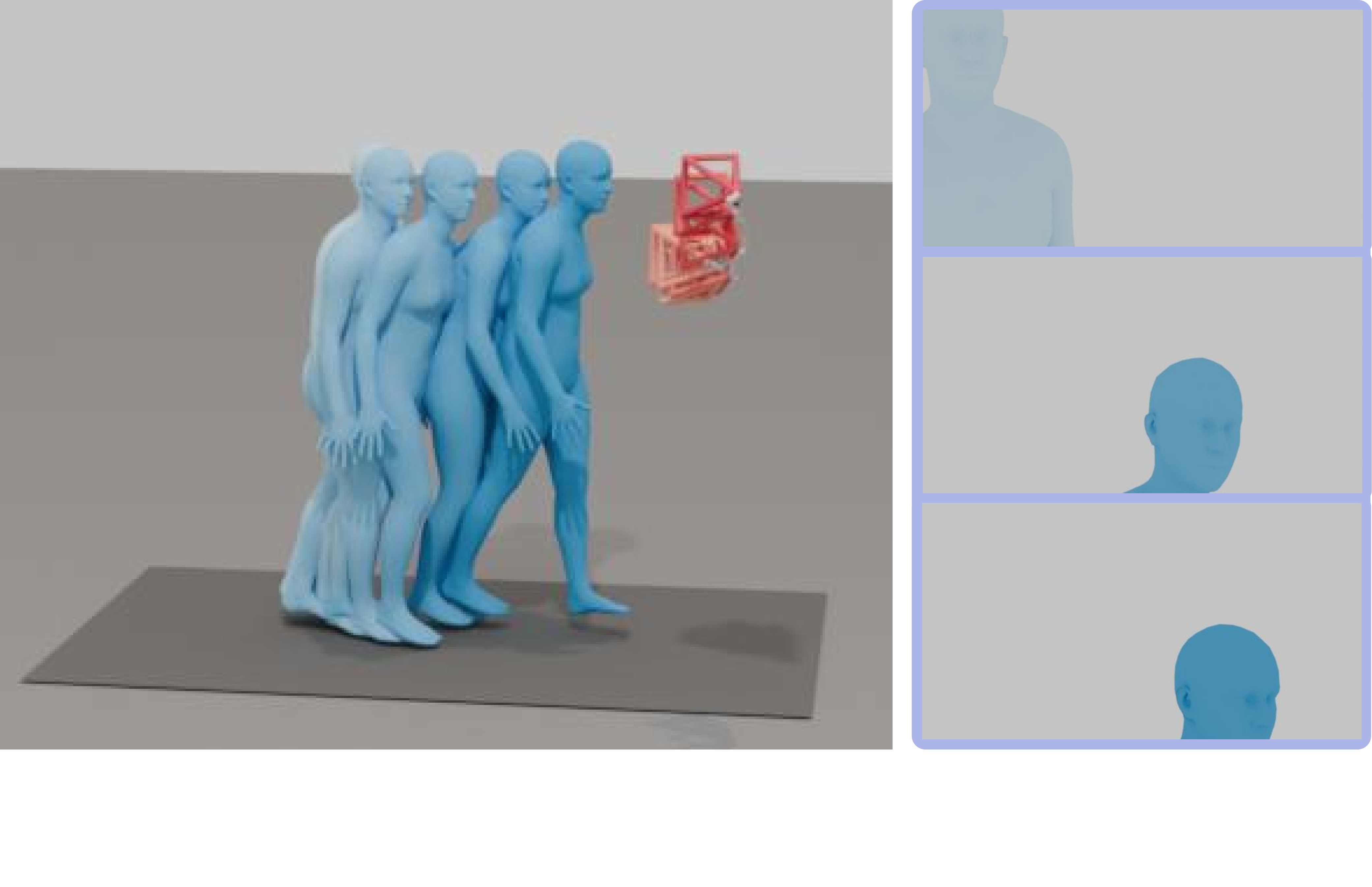}
    \put(0,5){\small \textbf{Human:} A person walks while turning head to right.}
    \put(0,0){\small \textbf{Camera:} The camera booms up.}
  \end{overpic}
  \captionof{figure}{\small \textbf{Example with MAR on the mixed set.}}
  \label{fig:qual-mixed-mar2}
\end{minipage}
\end{figure}

We show in \Cref{fig:qual-mixed-dit2} and \Cref{fig:qual-mixed-mar2} additional qualitative results on the \textit{mixed} subset. We also provide additional qualitative videos on our \href{https://pulp-motion.github.io/pulp-motion-gallery/}{project page} including 
\begin{itemize}
    \item \textbf{Generation examples}, comparing our method to baselines on both DiT and MAR architectures. We observe that our approach achieves more stable and consistent framing, reliably keeping the human on screen, whereas baselines either ignore the subject entirely or fail to maintain framing over the full sequence.
    \item \textbf{Dataset pipeline}, illustrating the extraction of camera, human, and textual information as well as the refinement step, which noticeably improves motion naturalness (e.g., converting sliding artifacts into realistic walking). 
    \item \textbf{Dataset examples}, highlighting the diversity of camera trajectories, human motions, and textual captions.
\end{itemize}

\subsubsection{Extra Comparison to the state of the art on pure dataset}

\begin{table}[ht]
    \caption{\small \textbf{State-of-the-art comparison on the pure subset.} We compare five baselines: human-conditioned camera generation \director{}~\cite{courant2024exceptional}, independent modality generation \indxy{}, dual-modality generation \jxy{}, triplet-modality generation \jxyz{}, and ReDi~\cite{kouzelis2025boosting}, along with our auxiliary sampling (\ours{}) method. Results are reported for DiT~\citep{peebles2023scalable} and MAR~\citep{li2024mar}. Superscript $\pm$ denotes the 95\% confidence interval over 10 samplings.}
    \label{tab:sota-pure}
    \centering
    \resizebox{\textwidth}{!}{%
    \begin{tabular}{l @{\hspace{4mm}} rrrr @{\hspace{4mm}} rrrr @{\hspace{4mm}} rr}
    \toprule
    \multirow{2}{*}{\centering \textbf{Methods}} & \multicolumn{2}{c}{\textbf{Framing}} & \multicolumn{4}{c}{\textbf{Human}} & \multicolumn{4}{c}{\textbf{Camera}} \\
    \cmidrule(r{4mm}){2-3} \cmidrule(r{4mm}){4-7} \cmidrule{8-11}
    & \multicolumn{1}{c}{$\text{FD}_{\text{framing}}\downarrow$}
    & \multicolumn{1}{c}{Out-rate $\downarrow$}
    & \multicolumn{1}{c}{$\text{FD}_{\text{TMR}}\downarrow$}
    & \multicolumn{1}{l}{TMR-Score $\uparrow$}
    & \multicolumn{1}{c}{R3 $\uparrow$}
    & \multicolumn{1}{c}{Coverage $\uparrow$}
    & \multicolumn{1}{c}{$\text{FD}_{\text{CLaTr}}\downarrow$}
    & \multicolumn{1}{c}{CLaTr-Score $\uparrow$}
    & \multicolumn{1}{c}{F1 $\uparrow$}
    & \multicolumn{1}{c}{Coverage $\uparrow$} \\
    \midrule
Ground-truth & 0.00 & 0.71 & 0.00 & 16.47 & 19.79 & 1.00 & 0.00 & 70.25 & 94.52 & 1.0 \\
Auto-encoder & 0.14 & 3.46 & 105.57 & 15.93 & 20.21 & 89.00 & 19.26 & 60.45 & 77.51 & 78.96 \\
    \midrule
    \multicolumn{10}{l}{DiT} \\
    \quad \director & 23.58$^{\pm 0.07}$ & 67.94$^{\pm 0.24}$ & - & - & - & - & 127.19$^{\pm 0.89}$ & 34.47$^{\pm 0.32}$ & 42.97$^{\pm 0.57}$ & 40.58$^{\pm 0.23}$ \\ 
    \quad \indxy{} & 10.24$^{\pm 0.08}$ & 41.70$^{\pm 0.40}$ & 384.31$^{\pm 0.62}$ & 23.72$^{\pm 0.10}$ & 20.63$^{\pm 0.35}$ & 10.46$^{\pm 0.18}$ & 86.06$^{\pm 0.38}$ & 57.74$^{\pm 0.29}$ & 75.53$^{\pm 0.23}$ & 31.83$^{\pm 0.21}$ \\
    \rowcolor{rowcolor!25}\quad \indxy{}+\ours{} (ours) & 7.88$^{\pm 0.07}$ & 36.03$^{\pm 0.36}$ & 443.65$^{\pm 0.89}$ & 24.67$^{\pm 0.09}$ & 21.41$^{\pm 0.35}$ & 8.63$^{\pm 0.19}$ & 77.78$^{\pm 0.57}$ & 62.75$^{\pm 0.31}$ & 83.00$^{\pm 0.35}$ & 29.53$^{\pm 0.28}$ \\
    \quad \jxy{} & 6.78$^{\pm 0.06}$ & 36.25$^{\pm 0.36}$ & 372.75$^{\pm 0.94}$ & 20.74$^{\pm 0.10}$ & 18.16$^{\pm 0.25}$ & 12.73$^{\pm 0.31}$ & 93.37$^{\pm 0.78}$ & 35.99$^{\pm 0.20}$ & 48.82$^{\pm 0.39}$ & 44.56$^{\pm 0.33}$ \\
    \quad \jxyz{} & 5.56$^{\pm 0.06}$ & 29.81$^{\pm 0.27}$ & 334.29$^{\pm 1.10}$ & 18.04$^{\pm 0.19}$ & 15.52$^{\pm 0.22}$ & 17.46$^{\pm 0.25}$ & 108.05$^{\pm 1.21}$ & 28.62$^{\pm 0.35}$ & 41.91$^{\pm 0.41}$ & 45.83$^{\pm 0.37}$ \\
    \rowcolor{rowcolor!25}\quad \jxyz{}+\ours{} (ours) & 4.66$^{\pm 0.04}$ & 23.61$^{\pm 0.36}$ & 438.38$^{\pm 1.44}$ & 19.47$^{\pm 0.10}$ & 4.85$^{\pm 0.19}$ & 14.78$^{\pm 0.21}$ & 83.65$^{\pm 1.29}$ & 30.80$^{\pm 0.31}$ & 41.30$^{\pm 0.59}$ & 47.06$^{\pm 0.34}$ \\
    \quad ReDi & 6.53$^{\pm 0.05}$ & 33.34$^{\pm 0.27}$ & 323.53$^{\pm 0.74}$ & 17.13$^{\pm 0.19}$ & 15.21$^{\pm 0.30}$ & 17.81$^{\pm 0.19}$ & 99.60$^{\pm 0.92}$ & 28.65$^{\pm 0.36}$ & 40.49$^{\pm 0.41}$ & 48.60$^{\pm 0.33}$ \\
    \rowcolor{rowcolor!50}\quad \jxy{}+\ours{} (ours) & 5.03$^{\pm 0.03}$ & 24.92$^{\pm 0.28}$ & 424.81$^{\pm 1.07}$ & 21.80$^{\pm 0.12}$ & 18.32$^{\pm 0.15}$ & 11.69$^{\pm 0.19}$ & 91.36$^{\pm 0.81}$ & 38.42$^{\pm 0.31}$ & 51.61$^{\pm 0.47}$ & 40.94$^{\pm 0.19}$ \\
    \midrule
    \multicolumn{10}{l}{MAR} \\
    \quad \indxy{} & 11.22$^{\pm 0.04}$ & 45.39$^{\pm 0.48}$ & 261.20$^{\pm 0.78}$ & 21.66$^{\pm 0.10}$ & 27.59$^{\pm 0.40}$ & 18.89$^{\pm 0.23}$ & 120.50$^{\pm 0.81}$ & 57.18$^{\pm 0.23}$ & 68.09$^{\pm 0.24}$ & 38.97$^{\pm 0.54}$ \\
    \rowcolor{rowcolor!25}\quad \indxy{}+\ours{} (ours) & 9.32$^{\pm 0.07}$ & 41.54$^{\pm 0.36}$ & 280.36$^{\pm 0.84}$ & 23.25$^{\pm 0.06}$ & 28.56$^{\pm 0.27}$ & 15.79$^{\pm 0.17}$ & 108.43$^{\pm 0.66}$ & 60.74$^{\pm 0.12}$ & 71.08$^{\pm 0.48}$ & 34.60$^{\pm 0.32}$ \\
    \quad \jxy{} &6.55$^{\pm 0.10}$ & 30.19$^{\pm 0.34}$ & 251.94$^{\pm 1.46}$ & 20.16$^{\pm 0.13}$ & 25.48$^{\pm 0.29}$ & 28.25$^{\pm 0.43}$ & 108.28$^{\pm 1.83}$ & 52.17$^{\pm 0.32}$ & 67.31$^{\pm 0.49}$ & 55.48$^{\pm 0.52}$ \\
    \quad \jxyz{} & 6.10$^{\pm 0.11}$ & 30.11$^{\pm 0.32}$ & 242.81$^{\pm 0.91}$ & 19.23$^{\pm 0.10}$ & 25.17$^{\pm 0.46}$ & 30.33$^{\pm 0.48}$ & 116.75$^{\pm 1.04}$ & 49.52$^{\pm 0.23}$ & 63.14$^{\pm 0.37}$ & 55.81$^{\pm 0.50}$ \\
    \rowcolor{rowcolor!25}\quad \jxyz{}+\ours{} (ours) & 4.30$^{\pm 0.05}$ & 23.48$^{\pm 0.28}$ & 262.34$^{\pm 0.61}$ & 21.08$^{\pm 0.14}$ & 10.30$^{\pm 0.30}$ & 19.26$^{\pm 0.34}$ & 108.98$^{\pm 0.59}$ & 52.36$^{\pm 0.38}$ & 66.16$^{\pm 0.47}$ & 49.72$^{\pm 0.46}$ \\
    \quad ReDi & 5.07$^{\pm 0.10}$ & 25.84$^{\pm 0.31}$ & 252.58$^{\pm 0.92}$ & 19.73$^{\pm 0.11}$ & 25.50$^{\pm 0.41}$ & 28.34$^{\pm 0.37}$ & 103.13$^{\pm 1.14}$ & 51.99$^{\pm 0.27}$ & 66.95$^{\pm 0.38}$ & 56.29$^{\pm 0.59}$ \\
    \rowcolor{rowcolor!50}\quad \jxy{}+\ours{} (ours) & 4.90$^{\pm 0.06}$ & 24.28$^{\pm 0.31}$ & 281.39$^{\pm 0.83}$ & 21.90$^{\pm 0.17}$ & 26.43$^{\pm 0.40}$ & 17.48$^{\pm 0.26}$ & 100.66$^{\pm 0.92}$ & 55.43$^{\pm 0.27}$ & 69.76$^{\pm 0.50}$ & 47.87$^{\pm 0.44}$ \\
    \bottomrule
    \end{tabular}}
    \label{supp:tab:sota-pure}
\end{table}
\begin{figure}[t]
\centering
\begin{minipage}{.49\textwidth}
  \centering
  \subcaptionbox{$\text{FD}_{\text{framing}}$-TMR-Score\label{fig:quant-pure-dit-clatr}}[.49\linewidth]{
    \input{fig/pure-comparisons/pure-dit-tmr}
  }
  \hfill
  \subcaptionbox{$\text{FD}_{\text{framing}}$-CLaTr-Score\label{fig:quant-pure-dit-tmr}}[.49\linewidth]{
    \input{fig/pure-comparisons/pure-dit-clatr}
    }

  \caption{\small \textbf{State-of-the-art comparison in DiT on the pure subset.} Trade-off between framing quality and modality-text alignment for textual guidance ranges from 5 to 12. The optimal region is at the bottom-right (low framing error, high alignment).}
  \label{fig:quant-pure-dit}
\end{minipage}%
\hfill
\begin{minipage}{.49\textwidth}
  \centering
  \subcaptionbox{$\text{FD}_{\text{framing}}$-TMR-Score\label{fig:quant-pure-mar-clatr}}[.49\linewidth]{
    \input{fig/pure-comparisons/pure-mar-tmr}
  }
  \hfill
  \subcaptionbox{$\text{FD}_{\text{framing}}$-CLaTr-Score\label{fig:quant-pure-mar-tmr}}[.49\linewidth]{
    \input{fig/pure-comparisons/pure-mar-clatr}
  }

  \caption{\small \textbf{State-of-the-art comparison in MAR on the pure subset.} Trade-off between framing quality and modality-text alignment for textual guidance values ranges from 2 to 5. The optimal region is at the bottom-right (low framing error, high alignment).}
  \label{fig:quant-pure-mar}
\end{minipage}
\end{figure}

\paragraph{Quantitative results.} 
In the main manuscript, we focused on the \textit{mixed} dataset. This section reports additional experiments trained and evaluated on the \textit{pure} subset.
Table~\ref{tab:sota-pure} reports a comparison of our auxiliary sampling (\ours{}) method against state-of-the-art baselines across both DiT and MAR architectures on the pure subset. We summarize our experimental observations as follows:
\newline \textbf{(i) Auxiliary sampling consistently improves framing (multimodal coherence).}
Applying \ours{} leads to systematic improvements in framing quality ($\text{FD}_{\text{framing}}$) and out-of-frame (Out-rate) rates across all baseline configurations and architectures.
For DiT, auxiliary sampling reduces $\text{FD}_{\text{framing}}$ from $11.24\!\to\!7.88$ for \indxy{}, from $6.78\!\to\!5.03$ when applied to \jxy{}. A similar trend holds for MAR, where $\text{FD}_{\text{framing}}$ decreases from $11.22\!\to\!9.32$ for \indxy{} and from $6.55\!\to\!4.90$ for \jxy{}.
Out rates follow the same pattern: for DiT, \ours{} reduces the out rate from $41,70\%\!\to\!36.03\%$ (\indxy{}) and from $36.25\%\!\to\!24,92\%$ (\jxy{}), while for MAR it decreases from $45.39\%\!\to\!41,54\%$ and $30.19\%\!\to\!24.28\%$, respectively. 
Overall, auxiliary sampling achieves the best $\text{FD}_{\text{framing}}$ and out rates across both architectures (DiT and MAR): showing that it is architecture-agnostic and consistently enhances framing quality, i.e. multimodal coherence, for all settings (independent and joint).
\newline \textbf{(ii) Conditioning on human motion alone is insufficient for strong framing.}
We next examine the human-conditioned camera generation baseline \director{}, where human trajectories are first generated using the same backbone as the independent setting (\indxy{}), and camera motion is subsequently conditioned on these synthesized humans. 
While this strategy is weaker than other baselines: for example, under DiT, \director{} yields a $\text{FD}_{\text{framing}}$ of $23.58$, compared to $7.88$ for \indxy{}+\ours{} and $5.03$ for \jxy{}+\ours{}. Its out rate is also higher with $67.94\%$ against $36.03\%$ and $24.92\%$, respectively. 
These results indicate that conditioning on human motion alone provides limited contextual information for accurate camera framing. 
In contrast, joint generation methods, especially when combined with \ours{}, consistently outperform \director{} and the independent baseline (\indxy{}), highlighting the importance of jointly modeling modalities during sampling.
\newline \textbf{(iii) Auxiliary sampling strengthens per-modality alignment while preserving quality.}
Beyond framing, \ours{} improves text–modality alignment across both human and camera dimensions. 
For human motion, auxiliary sampling increases TMR-Score from $20.74\!\to\!21.80$ (DiT, \jxy{}) and from $20.16\!\to\!21.90$ (MAR, \jxy{}), while also improving R3.
For camera motion, CLaTr-Score improves from $35.99\!\to\!38.42$ (DiT, \jxy{}) and from $52.17\!\to\!55.43$ (MAR, \jxy{}), along with gains in F1. 
Although modality generation quality metrics ($\text{FD}_{\text{TMR}}$ and $\text{FD}_{\text{CLaTr}}$) slightly increase, these changes are modest and do not outweigh the substantial gains in alignment and framing consistency. 
Overall, \ours{} achieves a favorable trade-off, improving multimodal coherence and per-modality alignment while maintaining strong generation quality across architectures.

Moreover, we compare our method with baselines for DiT and MAR in Figures~\ref{fig:quant-pure-dit} and~\ref{fig:quant-pure-mar}, showing the trade-off between framing quality ($\text{FD}_{\text{framing}}$) and modality-text alignment (TMR for human, CLaTr for camera) across different textual guidance values ($w_c$ in Equation~\ref{eq:epsilon-cfg-z}). The optimal point lies in the bottom-right corner of each plot (low $\text{FD}_{\text{framing}}$, high modality scores). 
Across both architectures and modalities, our auxiliary sampling (\ours{}) method achieves the best performance. It highlights its effectiveness in improving both framing quality and textual alignment on both architectures and for both modalities.

\begin{figure}[t]
\centering
\begin{minipage}{.49\textwidth}
  \centering
  \begin{overpic}[width=0.82\linewidth]{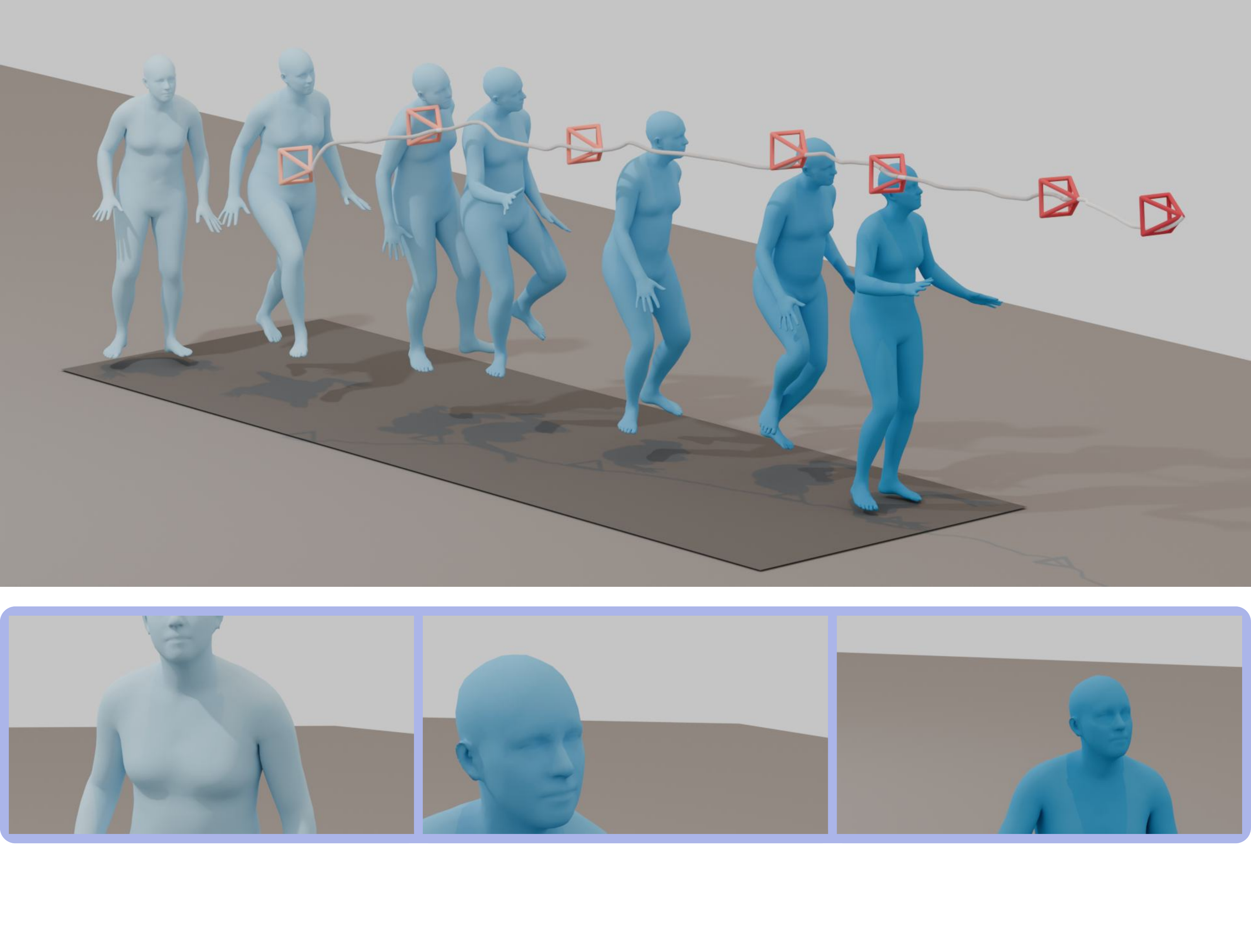}
    \put(0,4){\small \textbf{Human:} A person jumps.}
    \put(0,-2){\small \textbf{Camera:} The camera performs a pull out.}
  \end{overpic}
  \vspace{1.4mm}
  \captionof{figure}{\small \textbf{Example with DiT on the pure subset.}}
  \label{fig:qual-pure-dit}
\end{minipage}%
\hfill
\begin{minipage}{.49\textwidth}
  \centering
  \begin{overpic}[width=1.0\linewidth]{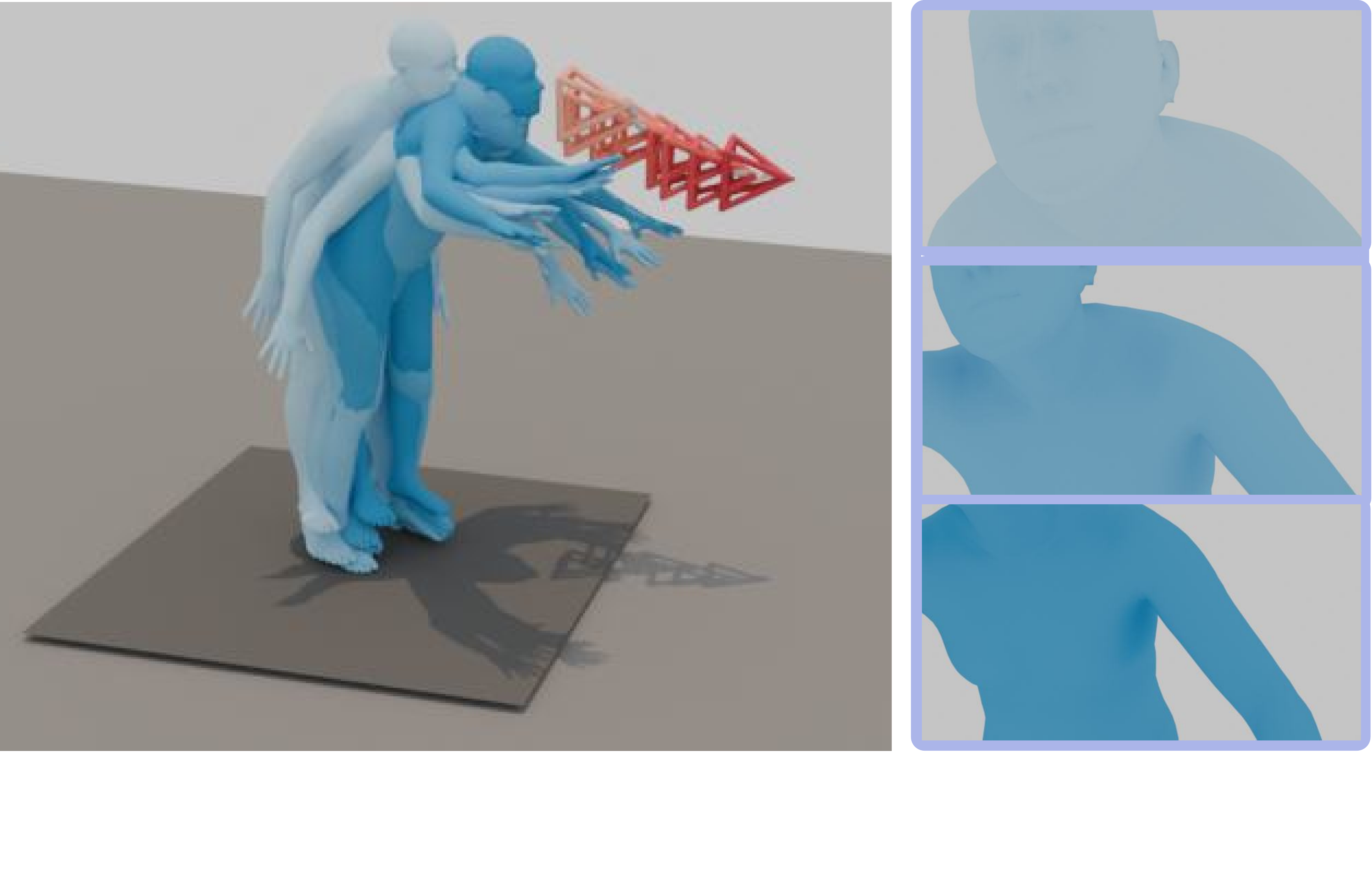}
    \put(0,5){\small \textbf{Human:} A person leans forward.}
    \put(0,0){\small \textbf{Camera:} The camera performs a pull out.}
  \end{overpic}
  \captionof{figure}{\small \textbf{Example with MAR on the pure subset.} }
  \label{fig:qual-pure-mar}
\end{minipage}
\end{figure}
\paragraph{Qualitative results.} 
Figures~\ref{fig:qual-pure-dit} and~\ref{fig:qual-pure-mar} present qualitative results with \ours{} sampling on the pure subset for DiT and MAR, respectively. 
In these examples, the human motion is accurately aligned with the prompts: in DiT, the person jumps as specified, and in MAR, the person leans forward. 
In the DiT example, the camera follows the person closely both vertically and laterally, maintaining proper on-screen framing, while in both cases the camera performs smooth pull-out motions. 
These examples further demonstrate that \ours{} generates human and camera behavior that faithfully follows the input prompts, producing precise motion and well-framed sequences across architectures.

\subsubsection{Ablation study on pure dataset}

\begin{table}[t]
    \caption{\small \textbf{Auxiliary guidance ablation on the pure subset.} We vary the auxiliary guidance weight $w_z$ to evaluate its effect on the framing, camera and human metrics. Results are reported for DiT~\citep{peebles2023scalable} and MAR~\citep{li2024mar}. Superscript $\pm$ denotes the 95\% confidence interval over 10 samplings.} 
    
    \label{tab:ablation-pure}
    \centering
    \resizebox{\textwidth}{!}{%
    \begin{tabular}{c @{\hspace{4mm}} rrrr @{\hspace{4mm}} rrrr @{\hspace{4mm}} rr}
    \toprule
    \multirow{2}{*}{\centering \textbf{$\bm{w_z}$}} & \multicolumn{2}{c}{\textbf{Framing}} & \multicolumn{4}{c}{\textbf{Human}} & \multicolumn{4}{c}{\textbf{Camera}} \\
    \cmidrule(r{4mm}){2-3} \cmidrule(r{4mm}){4-7} \cmidrule{8-11}
    & \multicolumn{1}{c}{$\text{FD}_{\text{framing}}\downarrow$}
    & \multicolumn{1}{c}{Out-rate $\downarrow$}
    & \multicolumn{1}{c}{$\text{FD}_{\text{TMR}}\downarrow$}
    & \multicolumn{1}{l}{TMR-Score $\uparrow$}
    & \multicolumn{1}{c}{R3 $\uparrow$}
    & \multicolumn{1}{c}{Coverage $\uparrow$}
    & \multicolumn{1}{c}{$\text{FD}_{\text{CLaTr}}\downarrow$}
    & \multicolumn{1}{c}{CLaTr-Score $\uparrow$}
    & \multicolumn{1}{c}{F1 $\uparrow$}
    & \multicolumn{1}{c}{Coverage $\uparrow$} \\
    \midrule
\multicolumn{10}{l}{DiT} \\
\quad $0.00$ & 6.78$^{\pm 0.06}$ & 36.25$^{\pm 0.36}$ & 372.75$^{\pm 0.94}$ & 20.74$^{\pm 0.10}$ & 18.16$^{\pm 0.25}$ & 12.73$^{\pm 0.31}$ & 93.37$^{\pm 0.78}$ & 35.99$^{\pm 0.20}$ & 48.82$^{\pm 0.39}$ & 44.56$^{\pm 0.33}$ \\
\rowcolor{rowcolor!50}\quad $0.25$ & 5.03$^{\pm 0.03}$ & 24.92$^{\pm 0.28}$ & 424.81$^{\pm 1.07}$ & 21.80$^{\pm 0.12}$ & 18.32$^{\pm 0.15}$ & 11.69$^{\pm 0.19}$ & 91.36$^{\pm 0.81}$ & 38.42$^{\pm 0.31}$ & 51.61$^{\pm 0.47}$ & 40.94$^{\pm 0.19}$ \\
\quad $0.50$ & 4.27$^{\pm 0.03}$ & 17.15$^{\pm 0.30}$ & 460.87$^{\pm 1.33}$ & 21.84$^{\pm 0.11}$ & 18.80$^{\pm 0.25}$ & 8.76$^{\pm 0.15}$ & 111.37$^{\pm 0.74}$ & 37.87$^{\pm 0.23}$ & 50.35$^{\pm 0.33}$ & 37.52$^{\pm 0.27}$ \\
\quad $0.75$ & 4.52$^{\pm 0.03}$ & 13.84$^{\pm 0.32}$ & 510.14$^{\pm 1.41}$ & 21.54$^{\pm 0.12}$ & 18.28$^{\pm 0.27}$ & 6.99$^{\pm 0.16}$ & 152.87$^{\pm 1.01}$ & 34.84$^{\pm 0.28}$ & 44.22$^{\pm 0.39}$ & 32.80$^{\pm 0.29}$ \\
\midrule
\multicolumn{10}{l}{MAR} \\
\quad $0.00$ & 6.55$^{\pm 0.10}$ & 30.19$^{\pm 0.34}$ & 251.94$^{\pm 1.46}$ & 20.16$^{\pm 0.13}$ & 25.48$^{\pm 0.29}$ & 28.25$^{\pm 0.43}$ & 108.28$^{\pm 1.83}$ & 52.17$^{\pm 0.32}$ & 67.31$^{\pm 0.49}$ & 55.48$^{\pm 0.52}$ \\
\rowcolor{rowcolor!50}\quad $0.50$ & 4.90$^{\pm 0.06}$ & 24.28$^{\pm 0.31}$ & 281.39$^{\pm 0.83}$ & 21.90$^{\pm 0.17}$ & 26.43$^{\pm 0.40}$ & 17.48$^{\pm 0.26}$ & 100.66$^{\pm 0.92}$ & 55.43$^{\pm 0.27}$ & 69.76$^{\pm 0.50}$ & 47.87$^{\pm 0.44}$ \\
\quad $1.00$ & 4.62$^{\pm 0.04}$ & 22.85$^{\pm 0.26}$ & 308.10$^{\pm 1.03}$ & 22.46$^{\pm 0.14}$ & 26.43$^{\pm 0.32}$ & 15.36$^{\pm 0.16}$ & 134.96$^{\pm 0.98}$ & 52.74$^{\pm 0.28}$ & 61.26$^{\pm 0.30}$ & 41.28$^{\pm 0.38}$ \\
\quad $1.50$ & 4.75$^{\pm 0.03}$ & 23.19$^{\pm 0.36}$ & 330.03$^{\pm 0.97}$ & 22.63$^{\pm 0.12}$ & 26.44$^{\pm 0.35}$ & 14.31$^{\pm 0.24}$ & 177.61$^{\pm 0.81}$ & 48.47$^{\pm 0.22}$ & 55.72$^{\pm 0.21}$ & 34.24$^{\pm 0.33}$ \\
\bottomrule
\end{tabular}}
\end{table}
\begin{figure}[H]
\centering
\begin{minipage}{.49\textwidth}
  \centering
  \captionsetup[subfigure]{skip=0.0pt}
  \subcaptionbox{\small $\text{FD}_{\text{framing}}$-TMR-Score\label{fig:ablation-pure-dit-tmr}}[.49\linewidth]{
    \input{fig/pure-ablations/pure-dit-tmr}
  }
  \subcaptionbox{\small $\text{FD}_{\text{framing}}$-CLaTr-Score\label{fig:ablation-pure-dit-clatr}}[.49\linewidth]{
    \input{fig/pure-ablations/pure-dit-clatr}
  }
  \caption{\small \textbf{Auxiliary guidance ablation in DiT.} Trade-off between framing quality and modality-text alignment for textual guidance ranges from 4 to 12. The optimal region is at the bottom-right (low framing error, high alignment).}
  \label{fig:ablation-pure-dit}
\end{minipage}%
\hfill
\begin{minipage}{.49\textwidth}
  \centering
  \captionsetup[subfigure]{skip=0.0pt}
  \subcaptionbox{\small $\text{FD}_{\text{framing}}$-TMR-Score\label{fig:ablation-pure-mar-tmr}}[.49\linewidth]{
    \input{fig/pure-ablations/pure-mar-tmr}
  }
  \subcaptionbox{\small $\text{FD}_{\text{framing}}$-CLaTr-Score\label{fig:ablation-pure-mar-clatr}}[.49\linewidth]{
    \input{fig/pure-ablations/pure-mar-clatr}
  }
  \caption{\small \textbf{Auxiliary guidance ablation in MAR.} Trade-off between framing quality and modality-text alignment for textual guidance ranges from 1 to 5. The optimal region is at the bottom-right (low framing error, high alignment).}
  \label{fig:ablation-pure-mar}
\end{minipage}
\end{figure}

Similarly to the manuscript, we ablate the auxiliary guidance weight $w_z$ (\Cref{eq:epsilon-cfg-z}) on both DiT and MAR on the \textit{pure} subset; results are shown in \Cref{tab:ablation-pure}.
We see that a moderate guidance weight improves framing and text–modality alignment. On DiT, increasing $w_z$ from $0.00$ to $0.25$ reduces $\text{FD}_{\text{framing}}$ $6.78\!\to\!5.03$ and Out-rate $36.25\!\to\!24.92$; on MAR, $w_z{=}0.50$ lowers them $6.55\!\to\!4.90$ and $30.19\!\to\!24.28$. 
\newline Pushing $w_z$ further continues to aid framing but degrades fidelity: $\text{FD}_{\text{TMR}}$ and $\text{FD}_{\text{CLaTr}}$ rise (DiT $424.81\!\to\!460.87$, MAR $281.39\!\to\!308.10$). 
\newline At high weights ($w_z{=}0.75$ for DiT, $1.50$ for MAR), the trend becomes unstable, with $\text{FD}_{\text{TMR}}$ spiking to $510.60$ and $\text{FD}_{\text{CLaTr}}$ to $177.61$.

We then illustrate Figures~\ref{fig:ablation-pure-dit} and~\ref{fig:ablation-pure-mar} for the trade-off between framing quality ($\text{FD}_{\text{framing}}$, lower is better) and text–modality alignment (TMR, CLaTr; higher is better) as the \ours{} guidance weight $w_z$ varies. The optimum lies near the bottom-right of each plot. Across both architectures, we see: (1) introducing guidance yields a large gain: $w_z{:}0\!\to\!0.25$ (DiT) and $0.50$ (MAR) shift points toward the bottom-right; (2) further increases, $0.50$ (DiT), $1.0$ (MAR), continue to improve framing but begin to reduce fidelity, reflected by larger markers (higher Fréchet distances); and (3) at very high weights, $0.75$ (DiT), $1.50$ (MAR), performance degrades on both axes.

\subsubsection{Ablation study on modality independence}
\label{sec:independence}

\begin{figure}[H]
  \centering
  \begin{minipage}{.49\textwidth}
    \centering
    \subcaptionbox{$\text{FD}_{\text{framing}}$-TMR-Score\label{fig:ablation-paired-mixed-dit-tmr}}[.49\linewidth]{
      \input{fig/mixed-paired/mixed-dit-tmr}
    }
    \subcaptionbox{$\text{FD}_{\text{framing}}$-CLaTr-Score\label{fig:ablation-paired-mixed-dit-clatr}}[.49\linewidth]{
      \input{fig/mixed-paired/mixed-dit-clatr}
    }
    \caption{\small \textbf{Independent modality ablation in DiT on mixed subset.} Trade-off between framing quality and modality-text alignment for textual guidance ranges from 4 to 12. The optimal region is at the bottom-right (low framing error, high alignment).}
    \label{fig:ablation-paired-mixed-dit}
  \end{minipage}%
  \hfill
  \begin{minipage}{.49\textwidth}
    \centering
    \subcaptionbox{$\text{FD}_{\text{framing}}$-TMR-Score\label{fig:ablation-paired-mixed-mar-tmr}}[.49\linewidth]{
      \input{fig/mixed-paired/mixed-mar-tmr}
    }
    \subcaptionbox{$\text{FD}_{\text{framing}}$-CLaTr-Score\label{fig:ablation-paired-mixed-mar-clatr}}[.49\linewidth]{
      \input{fig/mixed-paired/mixed-mar-clatr}
    }
    \caption{\small \textbf{Independent modality ablation in MAR on mixed subset.} Trade-off between framing quality and modality-text alignment for textual guidance ranges from 4 to 12. The optimal region is at the bottom-right (low framing error, high alignment).}
    \label{fig:ablation-paired-mixed-mar}
  \end{minipage}
  \end{figure}

\begin{figure}[H]
  \centering
  \begin{minipage}{.49\textwidth}
    \centering
    \subcaptionbox{$\text{FD}_{\text{framing}}$-TMR-Score\label{fig:ablation-paired-pure-dit-tmr}}[.49\linewidth]{
      \input{fig/pure-paired/pure-dit-tmr}
    }
    \subcaptionbox{$\text{FD}_{\text{framing}}$-CLaTr-Score\label{fig:ablation-paired-pure-dit-clatr}}[.49\linewidth]{
      \input{fig/pure-paired/pure-dit-clatr}
    }
    \caption{\small \textbf{Independent modality ablation in DiT on pure subset.} Trade-off between framing quality and modality-text alignment for textual guidance ranges from 4 to 12. The optimal region is at the bottom-right (low framing error, high alignment).}
    \label{fig:ablation-paired-pure-dit}
  \end{minipage}%
  \hfill
  \begin{minipage}{.49\textwidth}
    \centering
    \subcaptionbox{$\text{FD}_{\text{framing}}$-TMR-Score\label{fig:ablation-paired-pure-mar-tmr}}[.49\linewidth]{
      \input{fig/pure-paired/pure-mar-tmr}
    }
    \subcaptionbox{$\text{FD}_{\text{framing}}$-CLaTr-Score\label{fig:ablation-paired-pure-mar-clatr}}[.49\linewidth]{
      \input{fig/pure-paired/pure-mar-clatr}
    }
    \caption{\small \textbf{Independent modality ablation in MAR on pure subset.} Trade-off between framing quality and modality-text alignment for textual guidance ranges from 4 to 12. The optimal region is at the bottom-right (low framing error, high alignment).}
    \label{fig:ablation-paired-pure-mar}
  \end{minipage}
  \end{figure}

In this section, we analyze the influence of independent modality generation ($\rvx|\rvy$) versus dual-modality generation ($\rvx, \rvy$).
We illustrate the trade-off between framing quality and modality-text alignment for both DiT and MAR architectures in \Cref{fig:ablation-paired-mixed-dit} and \Cref{fig:ablation-paired-mixed-mar} for the \textit{mixed} subset, and in \Cref{fig:ablation-paired-pure-dit} and \Cref{fig:ablation-paired-pure-mar} for the \textit{pure} subset.

Across all settings, the same phenomena are consistently observed:
\begin{itemize}
    \item \textbf{The dual-modality generation setup \jxy{} tends to improve inter-modality alignment at the cost of lower modality-wise performance.} This is visible in the figures when comparing green vs. orange or red vs. brown curves: the dual-modality setting appears further to the left (worse modality-wise metrics) but lower on the vertical axis (better inter-modality alignment, framing).
    \item \textbf{In both independent and dual-modality cases, our auxiliary sampling (\ours{}) consistently enhances overall performance.} Comparing green vs. red and orange vs. brown curves, \ours{} shifts the points toward the bottom-right, closer to the optimal balance between framing quality and modality-text alignment.
\end{itemize}

\subsubsection{Qualitative visualization of auxiliary sampling influence}
\label{sec:influence}

\begin{figure}[h]
    \centering
    \begin{subfigure}[b]{0.32\textwidth}
        \centering
        \includegraphics[width=\textwidth]{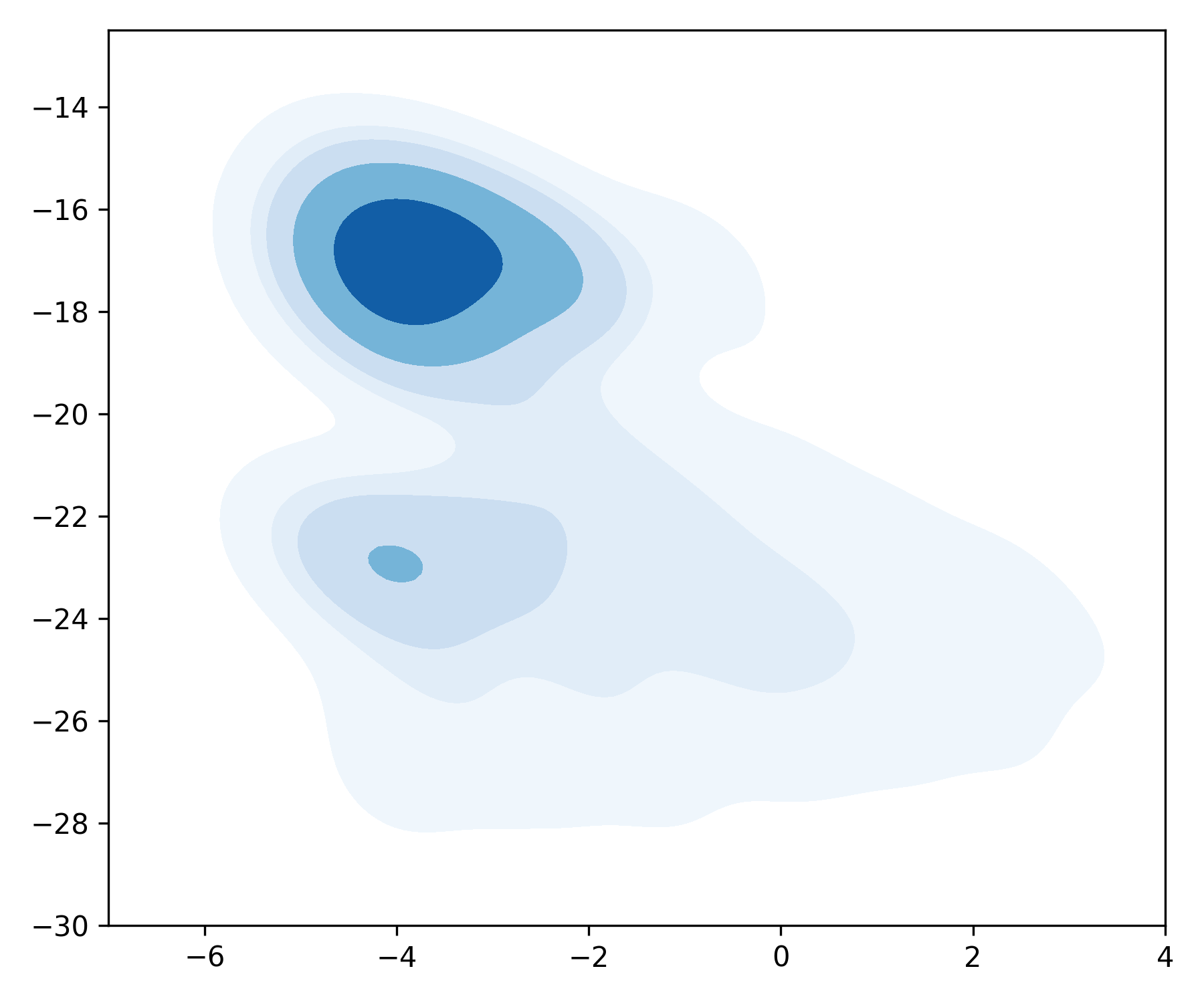}
        \caption{$w_z = 0.00$}
        \label{fig:wz0.00}
    \end{subfigure}
    \hfill
    \begin{subfigure}[b]{0.32\textwidth} 
        \centering
        \includegraphics[width=\textwidth]{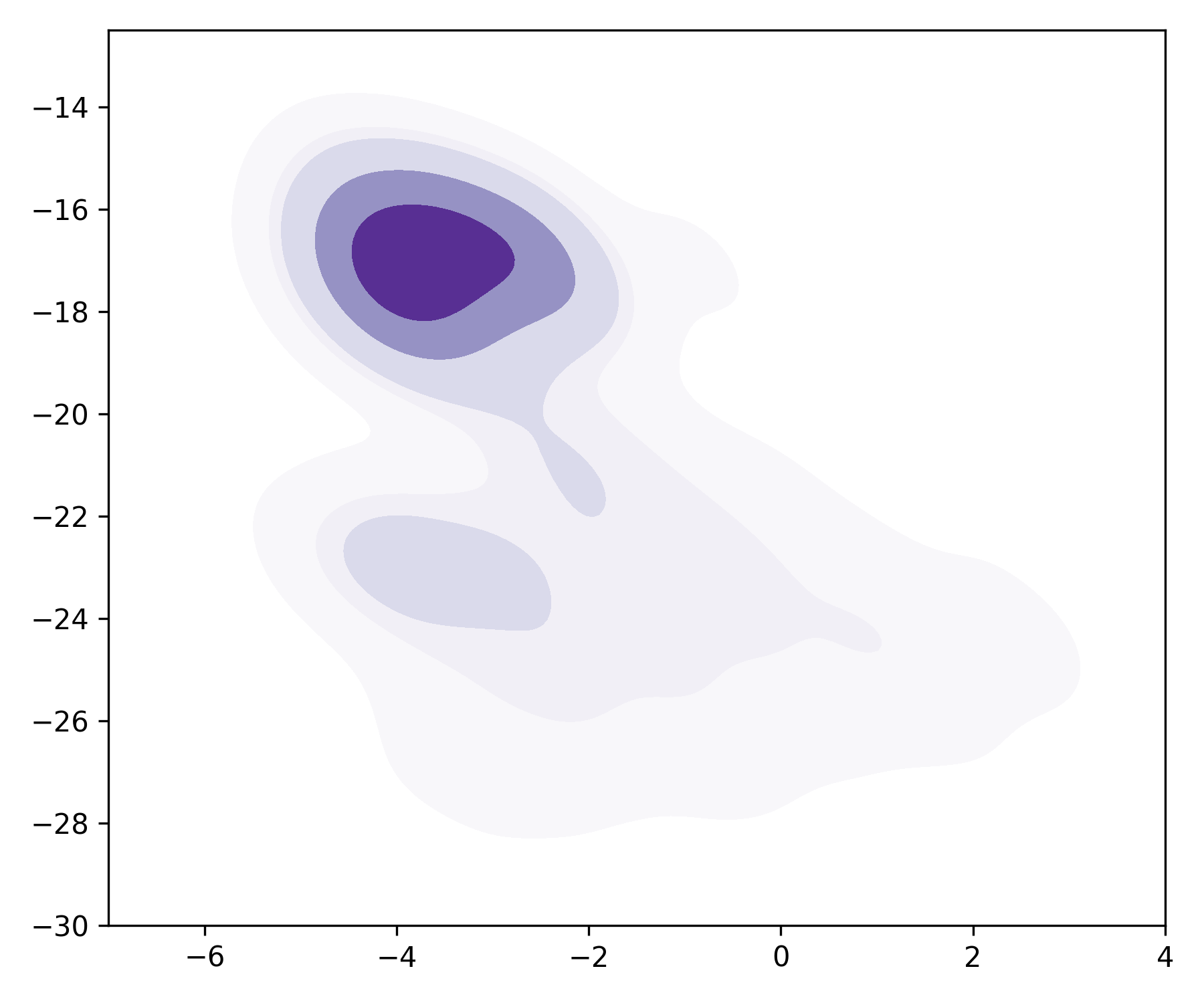}
        \caption{$w_z = 0.25$}
        \label{fig:wz0.25}
    \end{subfigure}
    \hfill 
    \begin{subfigure}[b]{0.32\textwidth} 
        \centering
        \includegraphics[width=\textwidth]{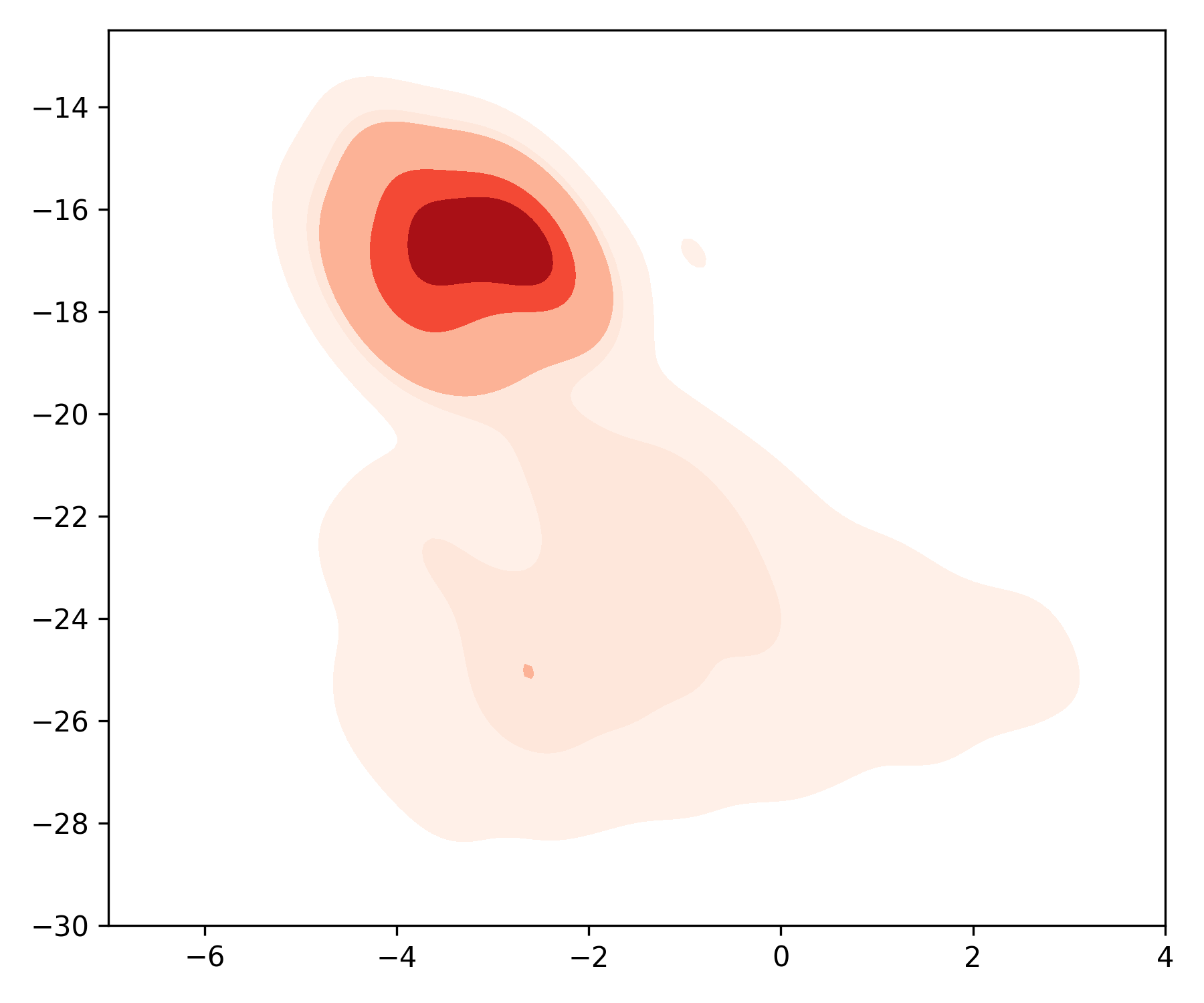}
        \caption{$w_z = 1.00$}
        \label{fig:wz1.00}
    \end{subfigure}

    \caption{Visualization of UMAP projection density of $2,000$ generated samples for different values of auxiliary sampling weight $w_z$.}
    \label{fig:umap-density} 
\end{figure}

\Cref{fig:umap-density} illustrates the evolution of UMAP projection density for a subset of generated samples as we vary the auxiliary sampling weight $w_z$. We analyze three settings: $w_z=0.00$ in \Cref{fig:wz0.00} (no auxiliary sampling, corresponding to the \jxy~baseline); $w_z=0.25$ in \Cref{fig:wz0.25} (the optimal value used for numerical comparisons); and $w_z=1.00$ in \Cref{fig:wz1.00} (an extreme value chosen to exaggerate the behavior).

Overall, we observe that auxiliary sampling shifts the probability mass towards a single mode. While \Cref{fig:wz0.00} exhibits bimodality with peaks near (-4,-22.5) and (-4,-16), increasing the weight to $w_z=1.00$ concentrates the density into a single mode around (-4,-16). 

We posit that this behavior mirrors the distribution tilting effect observed in guidance-based generation~\citep{karras2024guiding}, where stronger guidance pushes samples toward high-density regions of the conditional distribution, improving prompt adherence at the cost of reduced diversity. 
In our case, auxiliary sampling similarly strengthens cross-modal consistency (i.e. enhancing $\text{FD}_{\text{framing}}$ in \Cref{supp:tab:sota-pure}) while diminishing sample diversity (i.e. decreasing the modality coverage metrics \Cref{tab:sota-pure}).

\newpage

\end{document}